\documentclass[conference,10pt]{IEEEtran}

\usepackage{framed}
\usepackage{tcolorbox}
\usepackage{amsmath}
\usepackage{algorithm}
\usepackage{bbm}
\usepackage[noend]{algpseudocode}
\usepackage{bm}
\usepackage{stackrel}
\usepackage{subfig}
\usepackage{epsf}
\usepackage{amsmath,amssymb}
\usepackage{graphicx}
\usepackage{color}
\usepackage{cite}
\usepackage{multirow,tabularx}
\usepackage{ifthen}
\usepackage{epstopdf}
\input{epstopdf.sty}
\usepackage{times}

\input{epsf.sty}

\makeatletter
\def\BState{\State\hskip-\ALG@thistlm}
\makeatother

\newcommand{\hide}[1]{\ifthenelse{\boolean{false}}{#1}{}}

\newtheorem{theorem}{{\bf Theorem}}

\newtheorem{lemma}{{\bf Lemma}}

\newenvironment{proof}[1][Proof]{\begin{trivlist}
\item[\hskip \labelsep {\bfseries #1}]}{\end{trivlist}}

\newenvironment{definition}[1][Definition]{\begin{trivlist}
\item[\hskip \labelsep {\bfseries #1}]}{\end{trivlist}}

\newcommand{\qed}{\nobreak \ifvmode \relax \else
      \ifdim\lastskip<1.5em \hskip-\lastskip
      \hskip1.5em plus0em minus0.5em \fi \nobreak
      \vrule height0.75em width0.5em depth0.25em\fi}

\newcommand{\beq}{\begin{equation}}
\newcommand{\eeq}{\end{equation}}
\newcommand{\barr}{\begin{array}}
\newcommand{\earr}{\end{array}}

\newcommand{\benum}{\begin{enumerate}}
\newcommand{\eenum}{\end{enumerate}}

\newcommand{\bit}{\begin{itemize}}
\newcommand{\eit}{\end{itemize}}

\newcommand{\bc}{\begin{center}}
\newcommand{\ec}{\end{center}}

\newcommand{\bdes}{\begin{description}}
\newcommand{\edes}{\end{description}}

\newcommand{\bfig}{\begin{figure}}
\newcommand{\efig}{\end{figure}}

\newcommand{\bemq}{\begin{quote} \begin{em}}
\newcommand{\eemq}{\end{em} \end{quote}}

\newcommand{\bmp}{\begin{minipage}}
\newcommand{\emp}{\end{minipage}}

\newcommand{\apndx}[1]{Appendix~\ref{#1}}

\newcommand{\lemref}[1]{Lemma~\ref{#1}}
\newcommand{\thmref}[1]{Theorem~\ref{#1}}

\newcommand{\supth}{^{{\mathrm{th}}}}

\newcommand{\bsp}{\begin{slide*}}
\newcommand{\esp}{\end{slide*}}
\newcommand{\bsl}{\begin{slide}}
\newcommand{\esl}{\end{slide}}

\newcommand{\blem}{\begin{lemma}}
\newcommand{\elem}{\end{lemma}}
\newcommand{\bthm}{\begin{theorem}}
\newcommand{\ethm}{\end{theorem}}

\newcommand{\EX}[1]{\mathbb{E}\left[ #1 \right]} 
\newcommand{\pr}[1]{\mathbb{P}\left[ #1 \right]}

\IEEEoverridecommandlockouts

\begin{document}
\pdfoutput=1
\title{Age Optimal Information Gathering and Dissemination on Graphs}
\author{Vishrant Tripathi, Rajat Talak, and Eytan Modiano
\thanks{The authors are with the Laboratory for Information and Decision Systems (LIDS) at the Massachusetts Institute of Technology (MIT), Cambridge, MA. \texttt{ \{vishrant, talak, modiano\}@mit.edu.} This work was supported by NSF Grants AST-1547331, CNS-1713725, and CNS-1701964, and by Army Research Office (ARO) grant number W911NF-17-1-0508. A version of this work is to appear in IEEE Infocom 2019.}}

\IEEEaftertitletext{\vspace{-0.6\baselineskip}}

\maketitle
\begin{abstract}
We consider the problem of timely exchange of updates between a central station and a set of ground terminals $V$, via a mobile agent that traverses across the ground terminals along a mobility graph $G = (V, E)$. We design the trajectory of the mobile agent to minimize peak and average age of information (AoI), two newly proposed metrics for measuring timeliness of information.
We consider randomized trajectories, in which the mobile agent travels from terminal $i$ to terminal $j$ with probability $P_{i,j}$.
For the information gathering problem, we show that a randomized trajectory is peak age optimal and factor-$8\mathcal{H}$ average age optimal, where $\mathcal{H}$ is the mixing time of the randomized trajectory on the mobility graph $G$. We also show that the average age minimization problem is NP-hard. For the information dissemination problem, we prove that the same randomized trajectory is factor-$O(\mathcal{H})$ peak and average age optimal. Moreover, we propose an age-based trajectory, which utilizes information about current age at terminals, and show that it is factor-$2$ average age optimal in a symmetric setting.
\end{abstract}

\section{Introduction}
\label{sec:intro}
Many emerging applications depend on the collection and delivery of status updates between a set of ground terminals and a central terminal using mobile agents. Examples include: measuring traffic at road intersections~\cite{puri2005survey}, temperature, and pollution in cities~\cite{villa2016development}, ocean monitoring using underwater autonomous vehicles~\cite{paley2008cooperative}, and surveillance using UAVs~\cite{girard2004border}. All of these applications depend upon regular status updates, that are communicated in a timely manner, so as to keep the central terminal and the ground terminals updated with fresh information.

Age of Information (AoI) is a newly proposed metric that captures timeliness of the received information~\cite{kaul2012real, yin17_tit_update_or_wait, talak18_Mobihoc}. Unlike packet delay, AoI measures the lag in obtaining information at the destination node, and is therefore suited for applications involving gathering or dissemination of time sensitive updates. Age of information, at a destination, is defined as the time that elapsed since the last received information update was generated at the source. AoI, upon reception of a new update packet, drops to the time elapsed since generation of the packet, and grows linearly otherwise.

We consider the problem of AoI minimization in gathering and dissemination of information updates, between a set of ground terminals and a central terminal. The information updates can be as small as a single packet containing temperature information or a high fidelity image or a video file.
The ground terminals are equipped with low power transmitters, and a mobile agent is used to gather and disseminate information. 

The age or freshness of information gathered and disseminated depends on the trajectory of the mobile agent, whose mobility is constrained to a \emph{mobility graph} $G = (V,E)$. The mobile agent can move from ground terminal $i$ to ground terminal $j$ only if $(i,j) \in E$. This model can be used to capture the fact that the agent may not be able to move between any arbitrary locations due to topological limitations.

The problem of persistent monitoring in dynamic environments has been considered in \cite{cassandras2011optimal,lin2015optimal,smith2011persistent} using tools from optimal control. These works focus on minimizing uncertainty when source locations are time varying, rather than timely monitoring over a fixed set of locations. Minimizing delay in a similar setting with packets arriving randomly in space and time has been considered in \cite{celik2010dynamic}. 
There has also been work on trajectory control of a mobile agent for minimizing transmission energy in sensor networks \cite{ciullo2010minimizing}.

Closer to our work are \cite{alamdari2013min} and \cite{alamdari2014persistent}, in which some approximation trajectories to minimize maximum latency on metric graphs were proposed.  
In \cite{liu2018age}, the authors consider trajectory planning for a mobile agent to minimize AoI. They obtain the best permutation of nodes for the mobile agent to visit in sequence, given Euclidian distances between the nodes.
In our work, mobility is constrained by a general graph $G$, and we seek the optimal trajectory over the space of all trajectories allowed on this graph $G$, not just permutations of nodes.
To the best our knowledge, this is the first work to consider the AoI minimization on general mobility graphs $G$, and provide polynomial time approximation algorithms.

In the information gathering problem, we consider the design of trajectories for the mobile agent to minimizes peak and average age, two popular metrics of AoI.
We first consider the space of randomized trajectories, in which the mobile agent traverses edges according to a random walk on the mobility graph $G$. We show that a randomized trajectory is in fact peak age optimal, and that it can be obtained in polynomial time using the Metropolis-Hastings algorithm.
We then prove that solving for the average age optimal trajectory is NP-hard, in a symmetric setting, and
propose a heuristic randomized trajectory that is simultaneously peak age optimal and factor-$8{\mathcal{H}}$ average age optimal, where $\mathcal{H}$ is the mixing time of the randomized trajectory on $G$. The factor $\mathcal{H}$ can scale with the graph size, especially if the graph is not well connected. Thus, we propose an age-based trajectory, in which the mobile agent uses the current AoI to determine its motion, and show that it is factor-$2$ optimal in a symmetric setting.

In the information dissemination problem, the central terminal sends updates for each ground terminal via the mobile agent. The mobile agent queues these update packets in a first-come-first-serve (FCFS) queue, and delivers them to the respective ground terminal when the mobile agent reaches it. The FCFS queue assumption is motivated by uncontrollable MAC layer queues, where the generated updates get queued for transmission~\cite{2011SeCON_Kaul, talak18_Mobihoc}.
We, now, not only have to design the trajectory of the mobile agent, but also determine the optimal rate at which the central terminal generates information updates for each ground terminal.
We show that the peak age optimal randomized trajectory of the information gathering problem, along with a simple update generation rate, is at most a factor-$O(\mathcal{H})$ optimal, in both peak and average age. 
Also derived is an explicit formula for peak age of the discrete time Ber/G/1 queue with vacations, which may be of independent interest.

We describe the system model in Section~\ref{sec:model}. The information gathering and dissemination problems are studied in Section~\ref{sec:info_gathering} and Section~\ref{sec:info_diss}, respectively. We present simulation results in Section~\ref{sec:sim}, and conclude in Section~\ref{sec:conclusion}.

\section{System Model}
\label{sec:model}
We consider a central terminal that needs to communicate with a set of ground terminals $V$. The ground terminals are equipped with low power, low range radio communication devices, and cannot directly communicate with the central terminal, or with each other.
An autonomous mobile agent $m$, is used as a relay between the central terminal and the ground terminals, while moving across the geographical region where the ground terminals are spread.

The mobility of the agent is constrained by a \emph{mobility graph} $G = (V, E)$, where $m$ can travel from ground terminal $i$ to ground terminal $j$ only if $(i,j) \in E$.  The graph $G$, thus, constraints the set of allowable moves. We consider a time-slotted system, with slot duration normalized to unity. In the duration of a time-slot, the mobile agent stays at a ground terminal to gather or disseminate information, and moves to any of its neighbours in $G$ for the next time-slot. The mobility graph can be constructed from the limitations of a slot duration, distances between ground terminals, and speed of the mobile agent.

\begin{figure}
\centering
\includegraphics[width=0.85\linewidth]{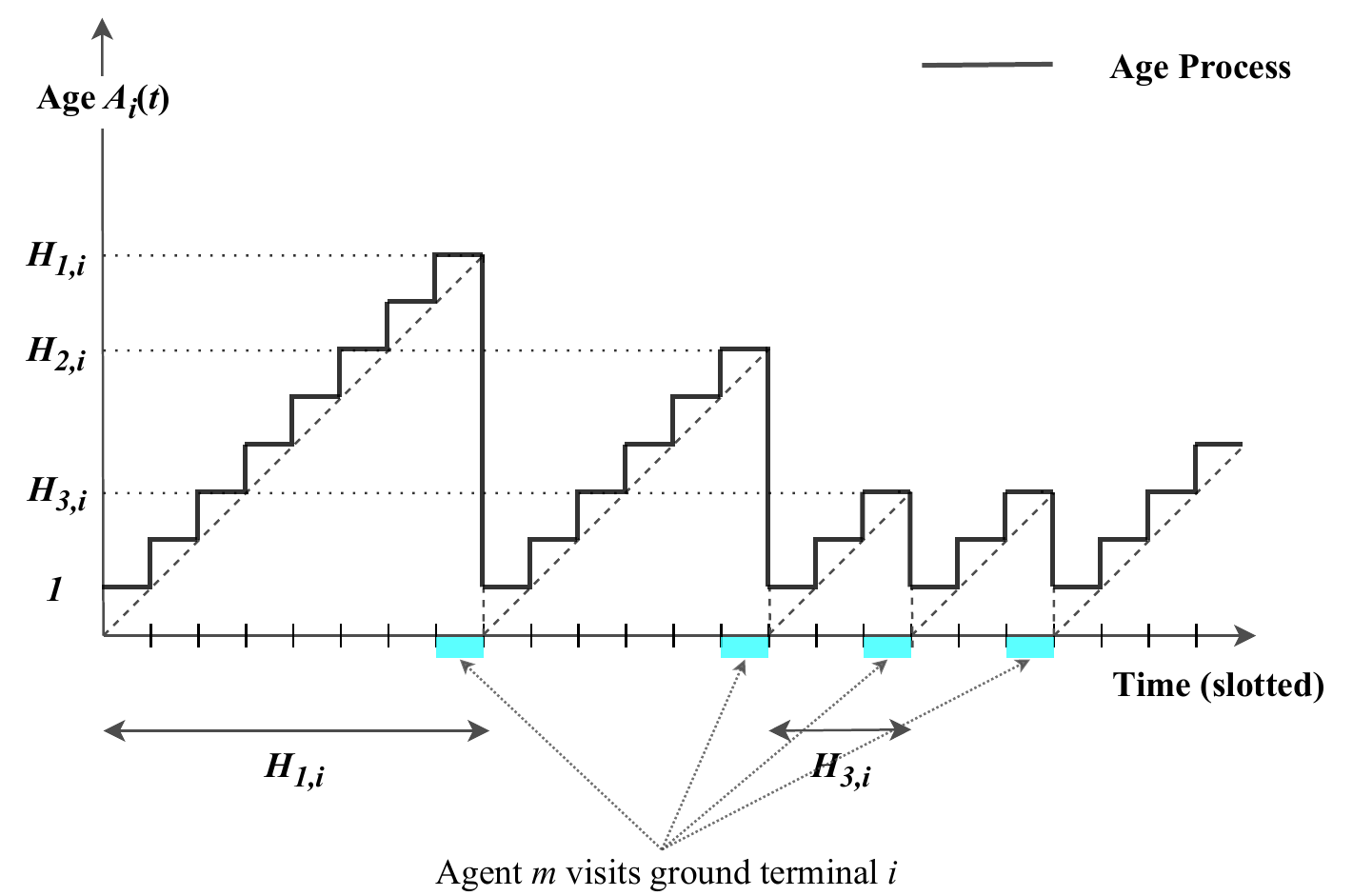}
\caption{Information gathering problem: time volution of age $A_{i}(t)$; $H_{k,i}$ is the $k\supth$ return time to terminal $i$. }
\label{figure:Fresh_Age}
\end{figure}

We consider two problems: \emph{information gathering} and \emph{information dissemination}. In the information gathering problem, every time the mobile agent reaches a ground terminal $i \in V$, the ground terminal sends a fresh update to the mobile agent, which is immediately relayed to the central terminal. The age $A_{i}(t)$, at the central terminal, for the ground terminal $i$ drops to $1$. When the mobile agent is not at the ground terminal $i$, the age $A_{i}(t)$ increases linearly. See Figure~\ref{figure:Fresh_Age}.
The evolution of $A_{i}(t)$ in the information gathering problem can be written as:
\begin{equation}
A_i(t+1) =
 \begin{cases}
      A_i(t)+1, & \text{if }m(t)\neq i \\
      1, &  \text{if }m(t) = i
 \end{cases}
\end{equation}
where $m(t)$ denotes the location of the mobile agent at time $t$. Note that the age evolution depends on the trajectory that the mobile agent follows on the mobility graph $G$.

In the information dissemination problem, the central terminal generates updates for each ground terminal. The generated updates are then transmitted to the mobile agent. 
The mobile agent queues updates received from the central terminal in a set of $V$ FCFS queues, one for each ground terminal. The mobile agent delivers the head-of-line update in queue $i$, to ground terminal $i$, when it reaches $i$. The central terminal has no control over the FCFS queues on the mobile agent, however, it can control the update generation rate $\lambda_i$, for each ground terminal $i$.

\begin{figure}
\centering
\includegraphics[width=0.85\linewidth]{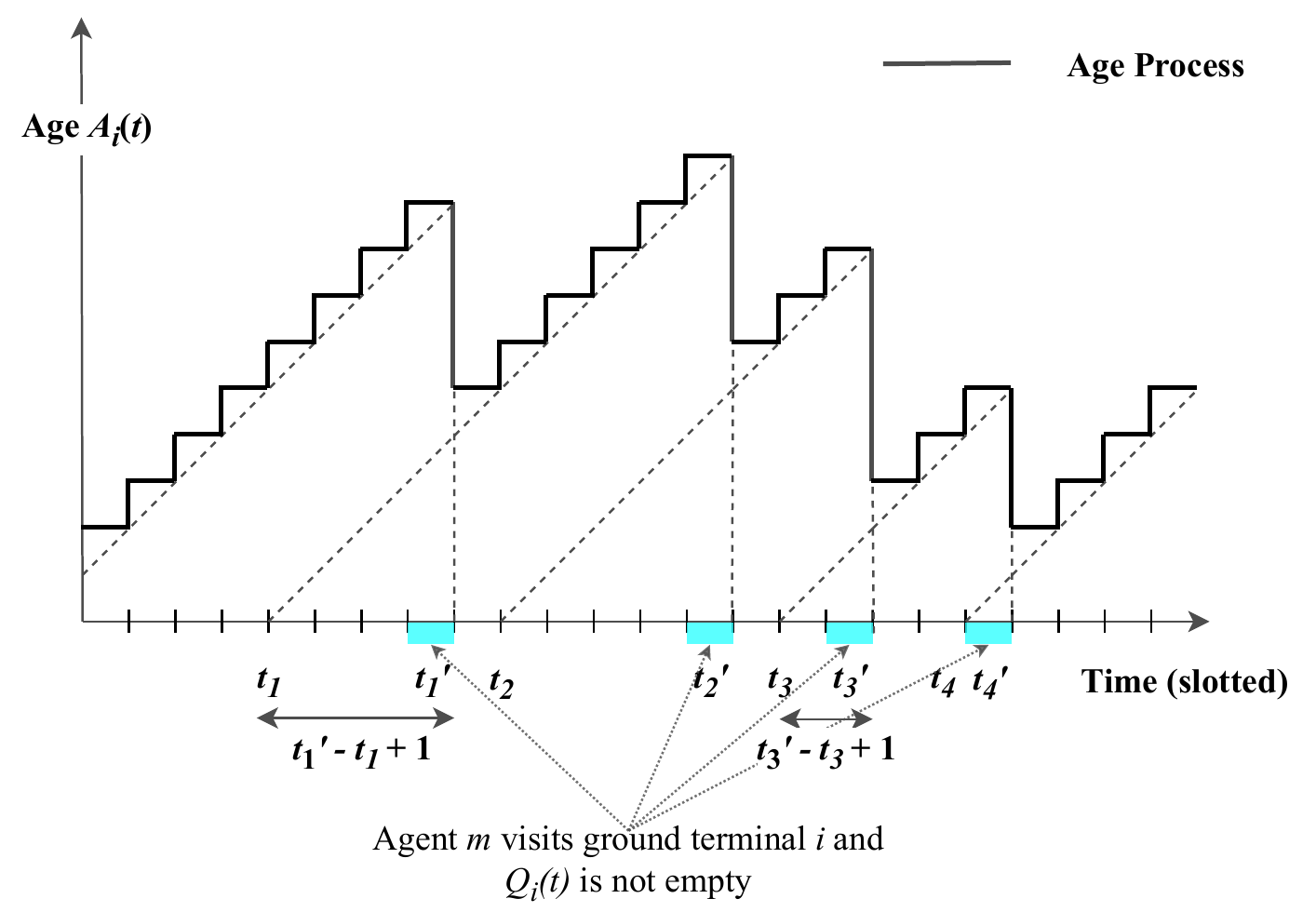}
\caption{Information dissemination problem: time evolution of age $A_{i}(t)$; $t_{k}, t_{k}'$ are the generation and reception times of the $k\supth$ status update for terminal $i$.}
\label{figure:FIFO_Age}
\end{figure}

The age $A_{i}(t)$, at the ground terminal $i$, increases by $1$ every time the mobile agent is not at $i$, or when it is at $i$ but has no updates to deliver. Otherwise, a successful delivery of the head-of-line update occurs in time slot $t$, and the age $A_{i}(t)$ drops to the age of the head-of-line update in queue $i$. See Figure~\ref{figure:FIFO_Age}. This evolution of age $A_{i}(t)$ can be written as:
\begin{equation}
\label{eq:aoi_evol_diss}
A_i(t+1) =
 \begin{cases}
      A_i(t)+1, & \text{if }m(t)\neq i \\
      A_i(t)+1, & \text{if }m(t) = i~\text{and}~\mathcal{Q}_i(t) = \emptyset  \\
      t - G_{i}(t) + 1, &  \text{if }m(t) = i~\text{and}~\mathcal{Q}_i(t) \neq \emptyset
 \end{cases},
\end{equation}
where $G_{i}(t)$ is the time of generation of the head of line packet in queue $i$, at time $t$, and $\mathcal{Q}_i(t)$ denotes the set of packets in the mobile agent's queue $i$ at time $t$.

\subsection{Age Metrics}
\label{sec:age_met}
AoI is an evolving function of time. We consider two time average metrics of AoI. Average age, for ground terminal $i$, is defined as the time averaged area under the age curve:
\begin{equation}
A_i^{\text{ave}} \triangleq \limsup_{T \rightarrow \infty} \frac{1}{T} \sum_{t = 1}^{T} A_i(t).
\end{equation}
In Figures~\ref{figure:Fresh_Age} and~\ref{figure:FIFO_Age}, we see that the age $A_i(t)$ peaks before a fresh update is delivered. In the information gathering case, a fresh update is delivered every time the mobile agent visits $i$, i.e. $m(t) = i$. Whereas, in the information dissemination case, a fresh update is delivered whenever $m(t) = i$ and the queue $\mathcal{Q}_i(t) \neq \emptyset$. The peak age $A_i^{\text{p}}$, for ground terminal $i$, defined as an average of all the peaks in the age evolution curve $A_i(t)$, can be written as
\begin{equation}
A_i^{\text{p}} \triangleq \limsup_{T \rightarrow \infty} \frac{ \sum\limits_{t = 1}^{t = T} A_i(t)\mathbbm{1}_{\{m(t) = i\}} }{ \sum\limits_{t = 1}^{t = T} \mathbbm{1}_{\{m(t) = i\}} },
\end{equation}
in the information gathering case and
\begin{equation}
A_i^{\text{p}} \triangleq \limsup_{T \rightarrow \infty} \frac{ \sum\limits_{t = 1}^{t = T} A_i(t)\mathbbm{1}_{\{m(t) = i, \mathcal{Q}_i(t) \neq \emptyset\}} }{ \sum\limits_{t = 1}^{t = T} \mathbbm{1}_{\{m(t) = i, \mathcal{Q}_i(t) \neq \emptyset\}} },
\end{equation}
in the information dissemination case.

We define the network peak and average age to be
\begin{equation}
A^{\text{p}} = \sum_{i \in V} w_i A^{\text{p}}_{i}~~~~\text{and}~~~~A^{\text{ave}} = \sum_{i \in V} w_i A^{\text{ave}}_{i},
\end{equation}
where $w_i > 0$ are weights representing the relative importance of a ground terminal $i$. Our goal is to minimize network peak and average age.

\subsection{Trajectory Space}
We use $\mathbb{T}$ to denote a reasonably large space of trajectories: 
\begin{equation}
\mathbb{T} = \left\{~\text{Trajectory}~\mathcal{T}~|~f_{i}(\mathcal{T})~\text{exists and is positive}~\forall~i \in V~\right\}, \nonumber 
\end{equation}
where $f_{i}(\mathcal{T})$ denotes the fraction of time-slots, the trajectory $\mathcal{T}$, is at ground terminal $i$:
\begin{equation}\label{eq:x}
f_{i}(\mathcal{T}) = \lim_{T \rightarrow \infty} \frac{1}{T}\sum_{t=1}^{T}\mathbbm{1}_{\{ m(t) = i \}}.
\end{equation}
For a trajectory $\mathcal{T} \in \mathbb{T}$, the limit~\eqref{eq:x} exists and is positive for all $i \in V$. This requirement is to ensure that the peak and average age are both finite and well defined.
 
Peak and average age depend on the trajectory $\mathcal{T} \in \mathbb{T}$. We use $A^{\text{p}}(\mathcal{T})$ and $A^{\text{ave}}(\mathcal{T})$ to denote network peak and average age, respectively, for $\mathcal{T} \in \mathbb{T}$.

\section{Information Gathering}
\label{sec:info_gathering}
In this section, we consider the problem of information gathering. We define optimal peak and average age to be
\begin{equation}\label{eq:opt_def}
A^{\text{p}\ast}_{\mathcal{G}} = \min_{\mathcal{T} \in \mathbb{T}} A^{\text{p}}(\mathcal{T}),~~~~\text{and}~~~~A^{\text{ave}\ast}_{\mathcal{G}} = \min_{\mathcal{T} \in \mathbb{T}} A^{\text{ave}}(\mathcal{T}),
\end{equation}
where $\mathbb{T}$ denotes the space of all trajectories for the mobile agent.

We first consider randomized trajectories, where the mobile agent moves according to a random walk on the mobility graph. We shall show that for peak age optimality, such randomized trajectories suffices. We then show that the average age optimization is NP-hard, and propose a heuristic randomized trajectory. In Section~\ref{sec:info_gath_age}, we propose an age-based trajectory for better average age performance.

\subsection{Randomized Trajectories}
\label{sec:info_gath_random}


We start by defining the class of randomized trajectories:
\begin{framed}
\begin{definition}
A trajectory $m(t)$, on mobility graph $G$, is said to be a \emph{randomized trajectory} if $m(t)$ is an irreducible Markov chain defined by a transition probability matrix $\mathbf{P}$:
\begin{equation}
\pr{m(t+1) = j | m(t) = i} = P_{i,j},
\end{equation}
for all $t$ and $i, j \in V$, where $P_{i,j} = 0$ for $(i,j) \notin E$.
\end{definition}
\end{framed}
For convenience, we shall refer to $m(t)$, defined above, as the randomized trajectory $\mathbf{P}$, where $\mathbf{P}$ to denote the matrix with entries $P_{i,j}$.
Note that $P_{i,j}$ is the probability that the mobile agent, when at ground terminal $i$, moves to ground terminal $j$ for the next time slot. The constraint: $P_{i,j} = 0$ for $(i,j) \notin E$, ensures that the randomized trajectory adheres to the mobility constraints defined by $G$.

We assume in the definition of a randomized trajectory $\mathbf{P}$, that $m(t)$ is an irreducible Markov chain over the state space $V$. This is desired, since the mobile agent has to traverse through all the nodes, repeatedly, for a positive fraction of time, or otherwise the resulting peak and average age would be unbounded.

For any randomized trajectory $\mathbf{P}$, we obtain explicit expressions for network peak and average age. We use the notation $A^{\text{p}}(\mathbf{P})$ and $A^{\text{ave}}(\mathbf{P})$ to show explicit dependence of peak and average age on the randomized trajectory $\mathbf{P}$.
\begin{framed}
\begin{theorem}
\label{thm:peakage}
The network peak and average age for a randomized trajectory $\mathbf{P}$ is given by
\begin{equation}
A^{\text{p}}(\mathbf{P}) = \sum_{i \in V} \frac{w_i}{\pi_i},~~~\text{and}~~~A^{\text{ave}}(\mathbf{P}) =  \sum_{i \in V} \frac{w_i z_{ii}}{\pi_i},
\end{equation}
where $\pi$ is the unique stationary distribution obtained by solving $\pi \mathbf{P} = \pi$ and $z_{ii}$ are diagonal elements of the matrix $Z \triangleq (I - \mathbf{P} + \Pi)^{-1}$, where $\Pi$ is an $n\times n$ matrix with entries $\Pi_{i,j} \triangleq \pi_j,  ~\forall i, j \in V$.
\end{theorem}
\end{framed}
\begin{IEEEproof}
The key step in proving the result above is to observe that the peak age of the ground terminal $i$, namely $A^{\text{p}}_{i}$, depends only on the mean of return times to terminal $i$; see Figure~\ref{figure:Fresh_Age}. Whereas, the average age $A^{\text{ave}}_{i}$ for $i$ depends on both, the mean and the variance, of return times to terminal $i$.

Given a randomized trajectory $\mathbf{P}$, the mean of return times to terminal $i$ is given by $\frac{1}{\pi_i}$, while the second moment of the return times is given by $\frac{-1}{\pi_i} + \frac{2z_{ii}}{\pi_i^2}$; see~\cite{aldous2002reversible}. Using this fact, we are able to obtain the explicit expressions for peak and average age.
A detailed proof is given in \apndx{pf:peakage}.
\end{IEEEproof}


\subsection{Peak Age Minimization}

We first formulate the peak age minimization problem over the space of randomized trajectories. We shall see that a peak age optimal randomized trajectory suffices for optimality over the space of all trajectories.

Using the results in Theorem~\ref{thm:peakage}, we can write the peak age minimization problem over the space of randomized trajectories as:
\begin{align}
\begin{aligned}
\label{peakopt1}
&\underset{\mathbf{P}, \mathbf{\pi}}{\text{Minimize}} &&\sum_{i \in V} \frac{w_i}{\pi_i}, \\
&\text{subject to} && P_{i,j} \geq 0, ~\forall (i,j), ~~\text{and}~~\mathbf{P}\mathbf{1} = \mathbf{1},\\
&&& \pi \mathbf{P} = \pi,~\mathbf{1}^{T}\pi = 1,~\text{and}~\pi_i \geq 0~\forall i  \\
&&& P_{i,j} = 0, ~\forall (i,j) \notin E,\\
&&& \mathbf{P}\text{ is irreducible.}
\end{aligned}
\end{align}
Note that $\mathbf{P}$ characterizes a randomized trajectory, while $\pi$ is the unique stationary distribution associated with it.

This problem is difficult to solve because the irreducibility constraint cannot be expressed in a simple, solvable manner. Further, relaxing the irreducibility constraint can yield a trivial solution like $\mathbf{P} = I$, which are neither irreducible nor anywhere close to optimal.


However, the problem~\eqref{peakopt1} can be transformed to finding an irreducible $\mathbf{P}$, with a given stationary distribution. This is a simpler problem and can be solved using the Metropolis-Hastings algorithm.

\begin{framed}
\begin{lemma}
\label{lem:stat_dist}
Let $\pi^{\ast}_i \triangleq \frac{\sqrt{w_i}}{\sum\limits_{j \in V} \sqrt{w_j}}$, for all $i \in V$, to be a distribution on $V$, and a randomized trajectory $\mathbf{P}$ satisfy $\pi^* \mathbf{P} = \pi^*$. Then, $(\pi^{\ast}, \mathbf{P})$ solves~\eqref{peakopt1}.
\end{lemma}
\end{framed}
\begin{IEEEproof}
See Appendix~\ref{pf:peakoptmh}.
\end{IEEEproof}

Lemma~\ref{lem:stat_dist} implies that a randomized trajectory $\mathbf{P}$, that satisfies $\pi^* \mathbf{P} = \pi^*$, is a peak age optimal, over the space of all randomized trajectories.
We now construct one such randomized trajectory: for $\pi^{\ast}$ given in Lemma~\ref{lem:stat_dist}, define a Metropolis-Hastings randomized trajectory $\mathbf{P}^{\text{mh}}$:
\begin{equation}
\label{eq:u1}
P^\text{mh}_{i,j} =
 \begin{cases}
      P^\text{rw}_{i,j}\min (1,\frac{\pi^*_j P^\text{rw}_{j,i}}{\pi^*_i P^\text{rw}_{i,j}}), & \text{if }i\neq j \text{ and }(i,j) \in E \\
      1- \sum\limits_{j : j \neq i} P_{i,j}^\text{mh}, &  \text{if } i = j
 \\
 0, & \text{otherwise}
\end{cases},
\end{equation}
where
\begin{equation}
\label{eq:u2}
P^\text{rw}_{i,j} =
 \begin{cases}
      \frac{1}{d_i}, & \text{if }i\neq j \text{ and }(i,j) \in E \\
      0, &  \text{otherwise }
 \end{cases}, \text{ } \forall i,j \in V,
\end{equation}
and $d_i$ equals the out degree of terminal $i$ in the mobility graph $G$. It is known that such a randomized trajectory $\mathbf{P}^{\text{mh}}$ satisfies $\pi^* \mathbf{P} = \pi^*$~\cite{aldous2002reversible}. We, therefore, have the following result.

\begin{framed}
\begin{theorem}
\label{thm:peak_random_tj} The Metropolis-Hastings randomized trajectory $\mathbf{P}^{\text{mh}}$ solves~\eqref{peakopt1}, i.e. it is peak age optimal over the space of all randomized trajectories.
\end{theorem}
\end{framed}
%

%

We considered randomized trajectories, where the mobile agent moves from terminal $i$ to $j$ with probability $P_{i,j}$. We now show that, for peak age optimality, such a randomization suffices.
\begin{framed}
\begin{theorem}
\label{thm:peak_opt_all}
The Metropolis-Hastings randomized trajectory $\mathbf{P}^{\text{mh}}$ is peak age optimal over the space of all trajectories $\mathbb{T}$, namely $A^{\text{p}\ast}(\mathbf{P}^{\text{mh}}) = A^{\text{p}\ast}_{\mathcal{G}}$.
\end{theorem}
\end{framed}
\begin{IEEEproof}
We establish a more general result. Namely, any randomized trajectory which satisfies $\pi^* \mathbf{P} = \pi^*$, where $\pi^{\ast}_i = \frac{\sqrt{w_i}}{\sum\limits_{j \in V} \sqrt{w_j}}$, is peak age optimal over the space of all trajectories:
\begin{equation}\nonumber
A^{\text{p}\ast}(\mathbf{P}) = A^{\text{p}\ast}_{\mathcal{G}}.
\end{equation}
To prove this, it suffices to argue that the peak age for any trajectory is lower bounded by $\sum_{i \in V} \frac{w_i}{\pi^{\ast}_i}$, where $\pi^{\ast}$ is as given in Theorem~\ref{thm:peak_random_tj}. We show this in~\apndx{pf:peaklb}.
\end{IEEEproof}
Thus, we are able to obtain a peak age optimal trajectory, namely $\mathbf{P}^{\text{mh}}$. Further, the matrix $\mathbf{P}^\text{mh}$ can be computed in polynomial time; in $O(|V|^2)$ time. Therefore, the peak age minimization problem is solved in polynomial time.

\subsection{Average Age Minimization}
We now consider the average age minimization problem. We first argue that in the symmetric setting, namely $w_i = 1~\forall~i \in V$,\footnote{The weights $w_i$ only measure relative significance of ground terminals. Thus, setting $w_i = 1~\forall~i \in V$ is equivalent to setting $w_i = w_j~\forall~i, j \in V$.} the average age minimization problem is NP-hard
\begin{framed}
\begin{theorem}
\label{thm:equalweight}
The problem of finding an average age optimal trajectory is NP-hard in the symmetric setting of $w_i = 1~\forall~i \in V$.
\end{theorem}
\end{framed}
\begin{IEEEproof}
See Appendix~\ref{pf:equalweight}. 
\end{IEEEproof}
Since solving the average age minimization problem is hard, we derive a lower bound on average age. 
Intuitively, if the mobility graph is better connected then it should yield a lower age. This is because a better connected mobility graph imposes fewer restrictions on mobility. The following result obtains a lower bound on network average age by comparing it with the network average age of a complete graph.
\begin{framed}
\begin{theorem}
\label{thm:lb}
For any trajectory $\mathcal{T} \in \mathbb{T}$, the network average age is lower bounded by
\begin{equation}
A^{\text{ave}}(\mathcal{T}) \geq \frac{1}{2} \sum_{i \in V} \bigg( \frac{w_i}{\pi^*_i} + w_i \bigg),
\end{equation}
where $\pi^{\ast}_{i} = \frac{\sqrt{w_i}}{\sum_{j \in V} \sqrt{w_j}}$ for all $i \in V$.
\end{theorem}
\end{framed}
\begin{proof}
See \apndx{pf:lb}.
\end{proof}
Note that the term $\sum_{i \in V} \frac{w_i}{\pi^*_i}$ is nothing but the optimal peak age $A^{\text{p}\ast}_{\mathcal{G}}$; see Theorem~\ref{thm:peak_opt_all}. Furthermore, the lower bound in Theorem~\ref{thm:lb} is independent of the trajectory $\mathcal{T}$. Therefore, we get
\begin{equation}
A^{\text{ave}\ast}_{\mathcal{G}} = \min_{\mathcal{T} \in \mathbb{T}} A^{\text{ave}}(\mathcal{T}) \geq A^\text{ave}_\text{LB} = \frac{1}{2}A^{\text{p}\ast}_{\mathcal{G}} + \frac{1}{2}\sum_{i \in V} w_i,
\end{equation}
where $\mathbb{T}$ is the space of all trajectories.
It must be noted that a similar result was derived in the case of link scheduling for age minimization in~\cite{talak18_Mobihoc}. The similarity of the result is rooted in the fact that the information gathering problem in the complete graph case is equivalent to the link scheduling problem in~\cite{talak18_Mobihoc}, in which at most one link can be activated simultaneously.

\subsubsection{A Heuristic Randomized Trajectory}
Motivated by the peak age optimality results of the previous section, we  restrict ourselves to the space of randomized trajectories, and propose a heuristic, called the \emph{fastest-mixing randomized trajectory}, and prove an average age performance bound for it.

Using the results in Theorem~\ref{thm:peakage}, the average age minimization problem over the space of randomized trajectories can be written as
\begin{align}
\begin{aligned}
\label{avgopt1}
&\underset{\mathbf{P}, \pi, \mathbf{Z}}{\text{Minimize}} && \sum_{i \in V}\frac{w_i z_{ii}}{\pi_i}, \\
&\text{subject to}
&& P_{i,j} \geq 0, ~\forall~(i,j),~~\text{and}~~\mathbf{P}\mathbf{1} = \mathbf{1},\\
&&& \pi\mathbf{P} = \pi,~\mathbf{1}^{T}\pi = 1,~\text{and}~\pi_i \geq 0~\forall i \\
&&& P_{i,j} = 0, ~\forall (i,j) \notin E, \\
&&& \mathbf{P}\text{ is irreducible}, \\
&&& \Pi_{i,j} = \pi_j~\forall~(i,j), \\
&&& Z = (I - \mathbf{P} + \Pi)^{-1}.
\end{aligned}
\end{align}
Here, $\mathbf{P}$ is the randomized trajectory and $\pi$ the unique stationary distribution corresponding to $\mathbf{P}$.
Solving~\eqref{avgopt1} can be computationally complex. Not only do we have the irreducibility constraint, but also a non-linear constraint in $Z = (I - \mathbf{P} + \Pi)^{-1}$.

We next upper bound the network average age, for any randomized trajectory $\mathbf{P}$ of the mobile agent. We first define mixing time for a randomized trajectory.

To do this, we first discuss the notion of stopping rules and stopping times in a Markov chain. A stopping rule is a rule that observes the walk on a Markov
chain and, at each step, decides whether or not to stop the walk based on the
walk so far. Stopping rules can make probabilistic decisions and therefore the time at which the walk stops, called the stopping time, is a random variable.
\begin{definition}[Mixing Time~\cite{king}] 
The hitting time from state distribution $\sigma_1$ to $\sigma_2$ on a Markov chain is
the minimum expected stopping time over all stopping rules that, beginning
at $\sigma_1$, stop in the exact distribution of $\sigma_2$. In other words, it is the expected
number of steps that the optimal stopping rule takes to move from $\sigma_1$ to $\sigma_2$.
This is denoted by $\mathcal{H}(\sigma_1,\sigma_2)$. The mixing time $\mathcal{H}$ of a
Markov chain $\mathbf{P}$ is then defined as
\begin{equation}
\mathcal{H} \triangleq \sup_{\sigma\in \bm{\Delta}(V)} \mathcal{H}(\sigma, \pi),
\end{equation}
where $\bm{\Delta}(V)$ is the collection of all distributions on $V$ and $\pi$ is the stationary distribution of $\mathbf{P}$. In other words, it is the expected time taken to reach stationarity using the optimal stopping rule and starting at the worst initial distribution.
\end{definition}
\begin{framed}
\begin{lemma}
\label{lem:approxlem}
The network average age for a randomized trajectory $\mathbf{P}$ is upper bounded by
\begin{equation}
A^{\text{ave}}(\mathbf{P}) = \sum_{i \in V} \frac{w_i z_{ii}}{\pi_i} \leq 4 \mathcal{H} A^{\text{p}}(\mathbf{P}) + \sum_{i \in V} w_i,
\end{equation}
where $\mathcal{H}$ denotes the mixing time of the randomized trajectory $\mathbf{P}$.
\end{lemma}
\end{framed}
\begin{IEEEproof}
First, we define the quantity \linebreak $\mathcal{Z} \triangleq \max\limits_i \sum\limits_j |z_{ij} - \pi_j|$, called the discrepancy of the randomized trajectory $\mathbf{P}$. This definition implies that $z_{ii} \leq \mathcal{Z} + \pi_i, ~\forall i \in V.$ Thus, we get the following upper bound:
\begin{equation}
\sum_{i \in V} \frac{w_i z_{ii}}{\pi_i} \leq \sum_{i \in V} \bigg(\frac{w_i \mathcal{Z}}{\pi_i}+w_i\bigg).
\end{equation}
However, from \cite{ailon2006clusters} we know that $\mathcal{Z} \leq 4\mathcal{H}$, where $\mathcal{H}$ is the mixing time of the randomized trajectory $\mathbf{P}$. Thus, we have the required result
\begin{equation}
\sum_{i \in V} \frac{w_i z_{ii}}{\pi_i} \leq \sum_{i \in V} \bigg(\frac{4w_i \mathcal{H}}{\pi_i}+w_i\bigg) = 4 \mathcal{H} A^{\text{p}}(\mathbf{P}) + \sum_{i \in V} w_i, \nonumber 
\end{equation}
where the last equality follows from Theorem~\ref{thm:peakage}.
\end{IEEEproof}
We use this relation and suggest the following heuristic for minimizing age: \emph{Find the fastest mixing randomized trajectory $\mathbf{P}$ on the mobility graph $G$ that minimizes peak age.}

From the proof of Theorem~\ref{thm:peak_opt_all}, we know that for a randomized trajectory $\mathbf{P}$ to be peak age optimal all we need is $\pi_i \propto \sqrt{w_i}$, where $\pi$ is the stationary distribution of $\mathbf{P}$. It, therefore, suffices to find $\mathbf{P}$ that satisfies $\pi_i \propto \sqrt{w_i}$, and simultaneously minimizes the mixing time $\mathcal{H}$. We call this the \emph{fastest-mixing randomized trajectory}, and use $\mathbf{P^*}$ to denote it.
The following result provides a way to obtain $\mathbf{P}^{\ast}$ by solving a convex program.
\begin{framed}
\begin{theorem}
\label{thm:convex_avg_age}
The fastest mixing randomized trajectory can be found by solving the following convex optimization problem: 
\begin{align}
\begin{aligned}
\label{op:avgh}
&\underset{\mathbf{P}}{\text{Minimize}} && \mu(\mathbf{P}) = || \mathbf{P} - \Pi^*||_2, \\
&\text{subject to}
&& P_{i,j} \geq 0, ~\forall (i,j), \\ &&&\mathbf{P}\mathbf{1} = \mathbf{1},\\
&&& \pi^* \mathbf{P} = \pi^*, ~~\Pi_{i,j}^{\ast} = \pi_i^{\ast}~\forall~i,j \in V,\\
&&& P_{i,j} = 0, \forall (i,j) \notin E.
\end{aligned}
\end{align}
Here $|| A ||_2$ denotes the spectral norm of matrix $A$ and $\pi_i^{\ast} = \frac{\sqrt{w_i}}{\sum_{j \in V} \sqrt{w_j}}, ~\forall i \in V$.
\end{theorem}
\end{framed}
\begin{IEEEproof}
See \apndx{pf:convex_avg_age}.
\end{IEEEproof}

This convex program~\eqref{op:avgh} finds a randomized trajectory $\mathbf{P}$ on $G$ that is closest to the stationary randomized walk $\Pi^*$, in the spectral norm sense. Also, $P^*$ is peak age optimal on graph $G$, since it satisfies $\pi_i^* \propto \sqrt{w_i}.$ Note that, the problem~\eqref{op:avgh} can be solved in polynomial time by converting it to a semi-definite program~\cite{boyd2004fastest}.

We now bound the average age performance of the fastest-mixing randomized trajectory.
\begin{framed}
\begin{theorem}
\label{thm:perform}
The network average age of the fastest-mixing randomized trajectory is at most $8\mathcal{H}$-factor away from the optimal average age:
\begin{equation}
\frac{A^{\mathrm{ave}}(\mathbf{P^*})}{A^{\text{ave}\ast}_{\mathcal{G}}} \leq 8 \mathcal{H},
\end{equation}
where $\mathcal{H}$ is the mixing time of $P^*$.
\end{theorem}
\end{framed}
\begin{proof}
See \apndx{pf:perform}.
\end{proof}
To see the usefulness of the fastest-mixing randomized trajectory, and Theorem~\ref{thm:perform}, consider a random geometric graph $\mathcal{G}(n,r)$. The graph consists of $n$ nodes spread over a unit square with a link between every two nodes that are within a distance $r$. If $v$ is the physical speed of the mobile agent, then $r$ must equal $v\tau$, where $\tau$ is the slot duration. We know that mixing time of $\mathcal{G}(n,r)$ is $O\left(\frac{\log n}{r^2}\right)$, and therefore, the fastest-mixing randomized trajectory would be at most $O\left(\frac{\log n}{v^2_{max} \tau^2}\right)$ factor optimal.
For highly connected graphs, such as Dirac graphs in which the degree of each node is at least $|V|/2$, we have constant factor of optimality; since the mixing times are $O(1). $~\cite{pak2002mixing} establishes a connection between the existence of long paths in graphs and their mixing times.

\subsection{Age-based Trajectories}
\label{sec:info_gath_age}
In the last two sub-sections, we proposed two randomized trajectories, namely $\mathbf{P}^{\text{mh}}$ and $\mathbf{P}^{\ast}$. Both were peak age optimal, while the latter was also factor-$\mathcal{H}$ average age optimal. We also noted that solving the average age problem is generally hard. We now propose an age-based trajectory which can be constant factor age optimal.

\begin{figure}
\centering
\includegraphics[width=0.65\linewidth]{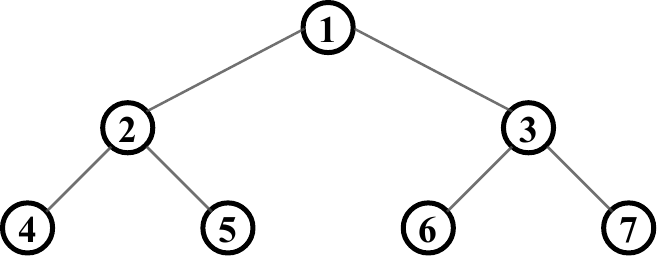}
\caption{Mobility graph restricted to a binary tree.}
\label{figure:tree}
\end{figure}

%
\begin{framed}
\begin{definition}[Age-based trajectory]
In every time slot, agent $m$ moves to the location that has the highest weighted function of $A_{i}(t)$. Specifically, if $m(t) = i$ then
\begin{equation}
m(t+1) = \arg\max_{j: (i,j) \in E} w_j g\left(A_{j}(t)\right),
\end{equation}
for all $i, j \in V$ and time $t$, where $g(\cdot)$ is an increasing function.
\end{definition}
\end{framed}
In the symmetric setting, where $w_i = 1~\forall~i \in V$, we observe that the age-based trajectory is a repeated depth-first traversal of the mobility graph $G$. This can be verified easily when the mobility graph is a tree. Consider the tree in Figure~\ref{figure:tree}, and assume that we start at the root node 1. The trajectory of the agent following the rule described above would be $1 \rightarrow 2 \rightarrow 4 \rightarrow 2 \rightarrow 5 \rightarrow 2 \rightarrow 1 \rightarrow 3 \rightarrow 6 \rightarrow 3 \rightarrow 7 \rightarrow 3 \rightarrow 1 ... $ This is precisely the depth-first traversal of the tree graph.

In the symmetric setting, where $w_i = 1~\forall~i \in V$, we now prove that the age-based trajectory is factor-2 optimal.
\begin{framed}
\begin{theorem}
\label{thm:age-based}
In the symmetric setting $w_i = 1~\forall~i \in V$, the network average age $A^{\text{ave}}$ for the age-based trajectory is bounded by
\begin{equation}
\frac{A^{\text{ave}}}{A^{\text{ave}\ast}_{\mathcal{G}}} \leq \frac{2|V|+1}{|V|+1} \leq 2,
\end{equation}
for any increasing function $g(\cdot)$.
\end{theorem}
\end{framed}
\begin{IEEEproof}
See Appendix~\ref{pf:thm:age-based}
\end{IEEEproof}
This age-based policy can be implemented in an online fashion if the mobile agent has access to age $A_{i}(t)$ of the neighboring terminals. The complexity of implementing this trajectory is then at most linear in the time-horizon and $|V|$. 

%

\section{Information Dissemination}
\label{sec:info_diss}
We now consider the information dissemination problem. The central terminal generates updates for every ground terminal $i$, at rate $\lambda_i$, according to a Bernoulli process. The generated updates for the ground terminal $i$ are sent to the mobile agent, which get queued in the $i$th FCFS queue. The mobile agent follows a trajectory $\mathcal{T}$, and delivers the head-of-line update in queue $i$ to terminal $i$, when it reaches it.

Our objective is to minimize the network peak age and average age over the space of update generation rates $\bm{\lambda}$ and all trajectories $\mathbb{T}$:
\begin{equation}\label{eq:opt_def}
A^{\text{p}\ast}_{\mathcal{D}} = \min_{\mathcal{T} \in \mathbb{T}, \bm{\lambda}} \sum_{i \in V} w_i A^{\text{p}}_{i},~~~~\text{and}~~~~A^{\text{ave}\ast}_{\mathcal{D}} = \min_{\mathcal{T} \in \mathbb{T}, \bm{\lambda}} \sum_{i \in V} w_i A^{\text{ave}}_{i},
\end{equation}
where $A^{\text{p}}_{i}$ denotes peak age and $A^{\text{ave}}_{i}$ denotes the average age of terminal $i$. Their evolution is given by~\eqref{eq:aoi_evol_diss}. For convenience, we have omitted their explicit dependence on $\mathcal{T} \in \mathbb{T}$ and $\bm{\lambda}$.

Motivated by the results for the information gathering problem, we consider randomized trajectories. Note that an arriving update in queue $i$ has service time equal to the inter-visit times to ground terminal $i$, provided the update arrived when the queue $i$ was not-empty; $\mathcal{Q}_i(t) \neq \emptyset$. However, when an update arrives to an empty queue $i$, the time to delivery is not the inter-visit time, and depends on the location of the mobile agent at the time of arrival. 

Since the analysis of age for such a queueing system may be difficult, we provide an upper bound, by comparing the the $i$th queue with a discrete time Ber/G/1 queue with vacations: whenever the $i$th queue is empty pretend that it goes on a vacation, with vacation times having the same distribution as inter-visit time; otherwise the service times for the queue are just inter-visit times.
Clearly, the age process of such a FCFS queue is an upper bound for the age process $A_{i}(t)$. 
Thus, we upper bound the peak age $A^{\text{p}}_i$ and average age $A^{\text{ave}}_{i}$, by the peak and average age of this Ber/G/1 queue with vacations. We first analyze peak and average age of a Ber/G/1 queue with vacations.

\subsection{Age for Ber/G/1 Queue with Vacations}
\label{sec:berG1}
Consider a discrete time FCFS Ber/G/1 queue with vacations, where an arrival occurs with probability $\lambda$, the service times $S$ are generally distributed with mean $\EX{S} = 1/\mu$, and the vacation times $V$ are also generally distributed.

We obtain an expression for the peak age of a discrete time Ber/G/1 queue with vacations, and a bound on average age.
\begin{framed}
\begin{lemma}
\label{lem:buff1}
The peak age for a discrete time FCFS Ber/G/1 queue with vacations is given by
\begin{equation}
\label{eq:buff_peak}
A^{\text{p}} = \frac{1}{\lambda} + \frac{1}{\mu} + \frac{\lambda \mathbb{E}[S^2]-\rho}{2(1-\rho)} + \frac{\EX{V^2}}{2\EX{V}}-\frac{1}{2},
\end{equation}
where
$\rho  = \frac{\lambda}{\mu}$, while the average age is upper-bounded by peak age, namely $A^{\text{ave}} \leq A^{\text{p}}$.
\end{lemma}
\end{framed}
\begin{IEEEproof}
The peak age for a FCFS queue is given by
\begin{equation}\label{eq:t1}
A^{\text{p}} = \EX{T + X},
\end{equation}
where $T$ denotes the time an update spends in the queue and $X$ is the inter-arrival time between two updates. Given that vacation times are distributed i.i.d according to random variable $V$, we have
\begin{equation}
\label{eq:t2}
\mathbb{E}[T] = \frac{\lambda \mathbb{E}[S^2]-\rho}{2(1-\rho)} + \frac{1}{\mu} +  \frac{\EX{V^2}}{2\EX{V}}-\frac{1}{2},
\end{equation}
where $S$ denotes the service time distribution. Substituting this and $\EX{X} = \frac{1}{\lambda}$ in~\eqref{eq:t1}, we obtain the expression for peak age. For the derivation of average system time $\EX{T}$, see~\apndx{pf:buff1}.

The upper-bound on average age directly follows from the observation that total time spent in the queue is negatively correlated with inter-arrival times. For details, see~\apndx{pf:buff1}.
\end{IEEEproof}

\subsection{Age Minimization Problem}
\label{sec:age_min_diss}
Using Lemma~\ref{lem:buff1}, we now obtain an upper-bound on both, network peak and average age, for a given randomized trajectory $\mathbf{P}$ and update generation rates $\bm{\lambda}$.
\begin{framed}
\begin{lemma}
\label{lem:buff2}
For a randomized trajectory $\mathbf{P}$ and packet generation rates $\bm{\lambda}$, the peak and average age for a ground terminal $i$ is upper-bounded by
\begin{equation}
A^{\text{UB}}_{i} = \frac{1}{\pi_i}\left[ 1 + z_{ii}+\frac{1}{\rho_i} + \frac{z_{ii}\rho_i}{1 - \rho_i}\right] - \frac{\rho_i}{1-\rho_i}-1,
\end{equation}
for all $i \in V$, where $\pi$ is the unique stationary distribution of $\mathbf{P}$, $Z = (I - \mathbf{P} + \Pi)^{-1}$, $\Pi$ is a matrix with all rows equal to the stationary distribution vector $\pi$, and $\rho_i \triangleq \frac{\lambda_i}{\pi_i}$.
\end{lemma}
\end{framed}
\begin{IEEEproof}
See Appendix~\ref{pf:buff2}.
\end{IEEEproof}

We propose a policy, i.e. a randomized trajectory $\mathbf{P}$ and update generation rate $\bm{\lambda}$, that minimizes the age upper-bound $A^{\text{UB}} = \sum_{i \in V} w_i A^{\text{UB}}_{i}$:
\begin{framed}
\begin{definition}
Separation Principle Policy

\begin{enumerate}
  \item Mobile agent follows the randomized trajectory $\mathbf{P}^{\ast}$ obtained by solving~\eqref{op:avgh}.
  \item Generate updates for the ground terminal $i$ at rate
        \begin{equation}
        \lambda_{i}^{\ast} = \frac{\pi^{\ast}_{i}}{1 + \sqrt{z^{\ast}_{ii} - \pi_{i}^{\ast}}},
        \end{equation}
        where $\pi^{\ast}_{i} = \frac{\sqrt{w_i}}{\sum_{j \in V} w_j}$ and $z_{ii}$ are diagonal elements of the matrix $Z = (I - \mathbf{P^*} + \Pi^*)^{-1}$.
\end{enumerate}
\end{definition}
\end{framed}
We call it the separation principle policy for two reasons. Firstly, $\mathbf{P}^{\ast}$ is the fastest-mixing randomized trajectory, which we proposed for minimizing average age in the information gathering problem. Secondly, the update generation rate for the ground terminal $i$, depends only on $z_{ii}$ and $\pi_i$, which are functions of the first and second moments of the return times to terminal $i$ under trajectory $\mathbf{P}^{\ast}$:
\begin{equation}\nonumber
\EX{H_{i}} = \frac{1}{\pi_i}~\text{and}~\EX{H_{i}^{2}} = - \frac{1}{\pi_i} + \frac{2z_{ii}}{\pi_i},
\end{equation}
where $H_{i}$ denotes the return time to terminal $i$, starting from $i$, under the fastest mixing randomized trajectory $\mathbf{P}^{\ast}$.
We now bound the performance of this separation principle policy.
\begin{framed}
\begin{theorem}
\label{thm:buff_perform}
The peak and average age of the separation principle policy is bounded by
\begin{equation}\nonumber
\frac{A^{\text{p}}}{A^{\text{p}\ast}_{\mathcal{D}}} \leq 4{\mathcal{H}} + 4\sqrt{\mathcal{H}} + 2~\text{and}~\frac{A^{\text{ave}}}{A^{\text{ave}\ast}_{\mathcal{D}}} \leq 8{\mathcal{H}}+ 8\sqrt{\mathcal{H}} + 4,
\end{equation}
where $\mathcal{H}$ is the mixing time of the randomized trajectory $\mathbf{P}^{\ast}$.
\end{theorem}
\end{framed}
\begin{IEEEproof}
See Appendix~\ref{pf:buff_perform}.
\end{IEEEproof}

The separation principle policy is factor $O({\mathcal{H}})$ peak age and average age optimal. It is worthwhile to note that a similar separation principle policy was established in a completely different setting of scheduling links for age minimization in~\cite{talak18_Mobihoc}. Theorem~\ref{thm:buff_perform} generalizes that result to a graph.

\section{Simulation Results}
\label{sec:sim}
\begin{figure*}
\centering
\subfloat[]{\includegraphics[width=61mm]{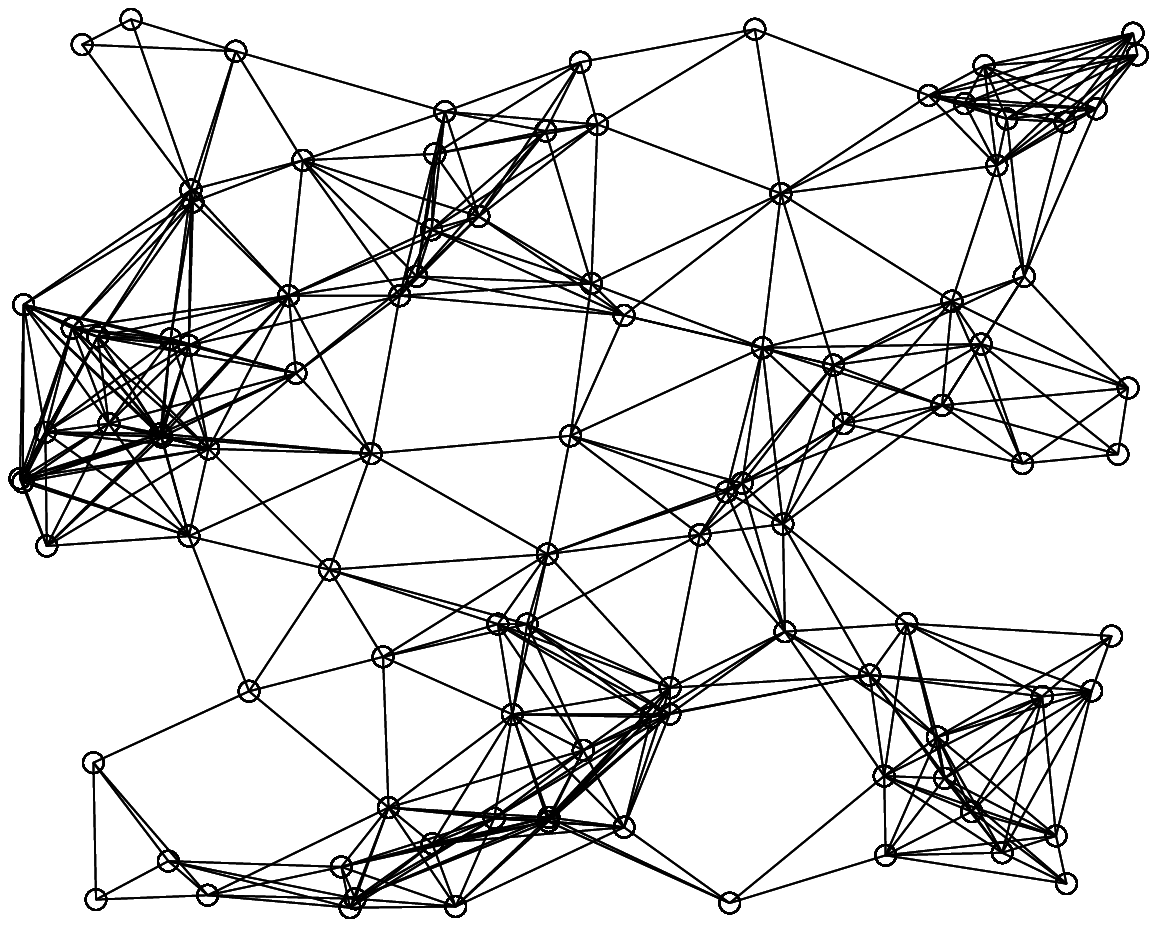}}
\subfloat[]{\includegraphics[width=60mm]{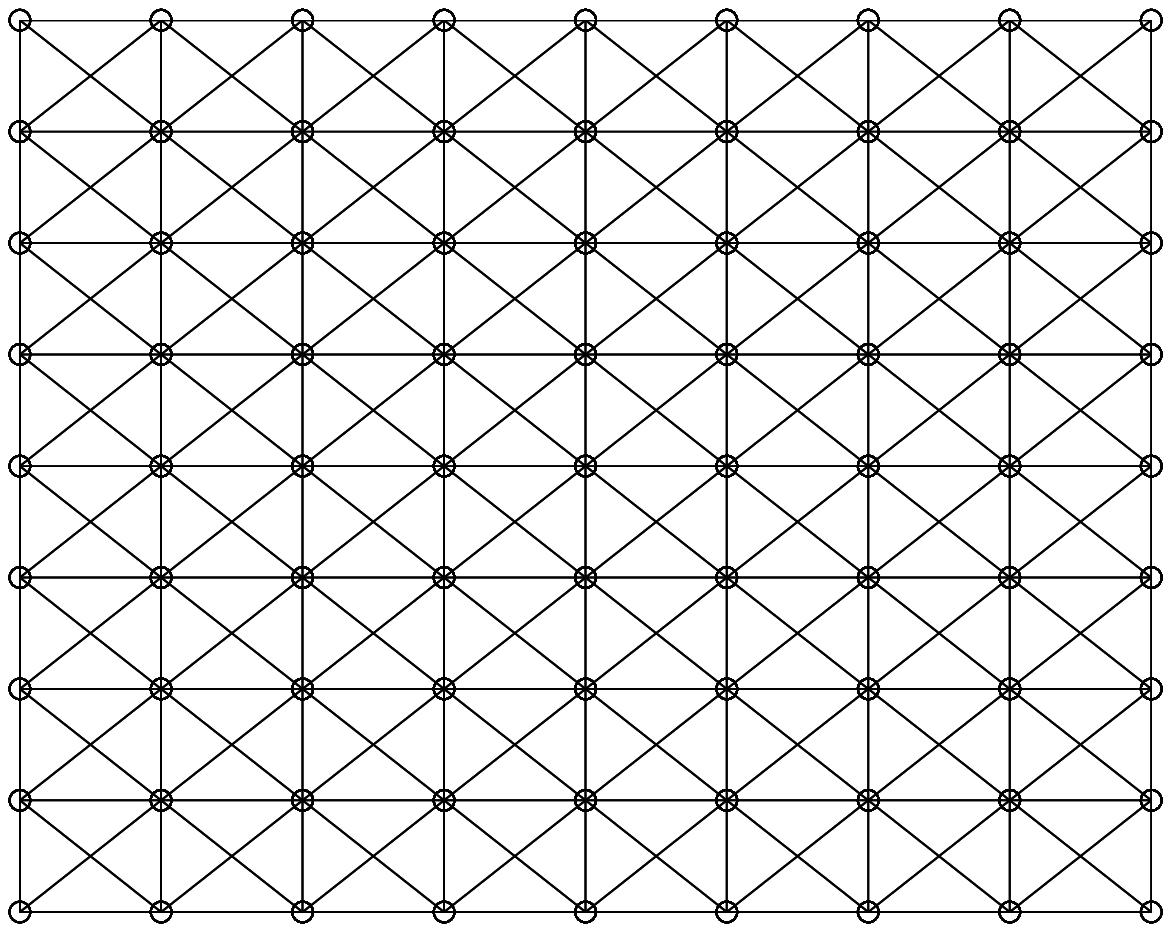}}
\subfloat[]{\includegraphics[width=53mm]{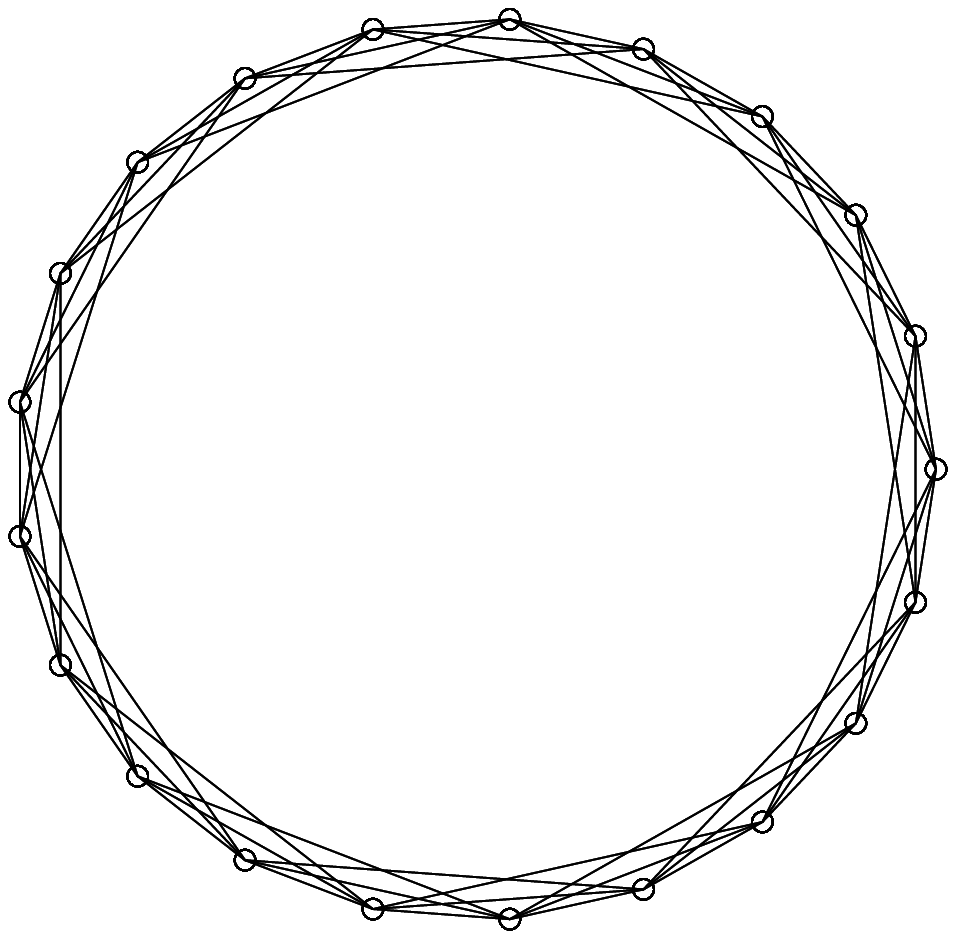}}
\caption{(a) A random geometric graph with 100 nodes, (b) A grid graph with 81 nodes and diagonal edges, and (c) A 3-connected ring or cycle graph with 21 nodes.}
\label{figure:graphs}
\end{figure*}
We test the performance of our proposed trajectories on three different kinds of mobility graphs: random geometric graphs $\mathcal{G}(n, \frac{2}{\sqrt{n}})$,\footnote{Setting $r = \frac{2}{\sqrt{n}}$ for random geometric graphs ensures connectivity w.h.p.} grid graphs with diagonal edges, and 3-connected ring or cycle graphs; see Figure~\ref{figure:graphs}. We use $n$ to denote the number of ground terminals, namely $n = |V|$. For the age-based policy, we set the function $g(a) = a^2 + a$, inspired by the index based policies in \cite{talak18_Mobihoc}. Link weights are picked uniformly at random from the interval $(1,2]$ in an independent manner. We run our simulations for a total of $50000$ time-slots, to get a good estimate of the peak and average age.

\begin{figure}
\centering
\includegraphics[width=\linewidth]{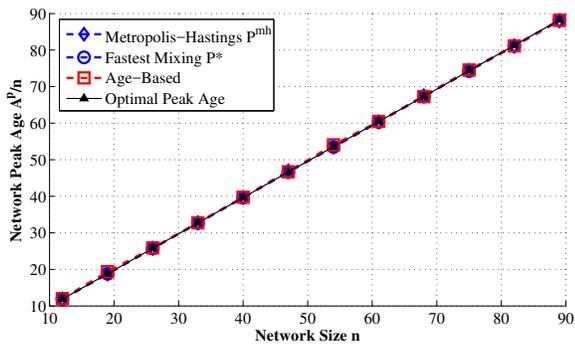}
\caption{Information gathering problem in $\mathcal{G}(n,2/\sqrt{n})$: network peak age as a function of network size $n$ for several proposed trajectories of the mobile agent.}
\label{figure:random_geometric}
\end{figure}
We first consider the information gathering problem, and plot peak and average age for all the proposed trajectories of the mobile agent: the Metropolis-Hastings randomized trajectory $\mathbf{P}^{\text{mh}}$, fastest mixing randomized trajectory $\mathbf{P}^{\ast}$, and age-based trajectory.
Figure~\ref{figure:random_geometric} plots peak age as a function of network size $n$ for the random geometric graph $\mathcal{G}\left(n,2/\sqrt{n}\right)$. We observe that the peak age for all the three proposed trajectories match. We know from Theorems~\ref{thm:peak_opt_all} and~\ref{thm:convex_avg_age} that that the two randomized trajectories, namely, the Metropolis-Hastings randomized trajectory $\mathbf{P}^{\text{mh}}$ and the fastest mixing randomized trajectory $\mathbf{P}^{\ast}$, are both peak age optimal. Figure~\ref{figure:random_geometric}, therefore, suggests that even the age-based trajectory for the mobile agent is peak age optimal.

\begin{figure}
\centering
\includegraphics[width=\linewidth]{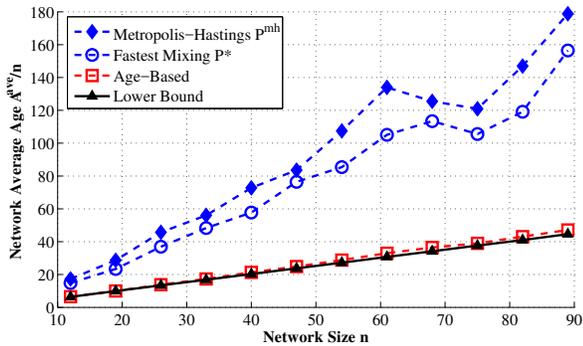}
\caption{Information gathering problem in $\mathcal{G}(n,2/\sqrt{n})$: network average age as a function of network size $n$ for several proposed trajectories of the mobile agent.}
\label{figure:random_geometric_avg}
\end{figure}
In Figure~\ref{figure:random_geometric_avg} we plot the average age performance of the proposed trajectories, as a function of network size $n$. Also plotted is the lower bound for average age derived in Theorem~\ref{thm:lb}. We see that the age-based policy is nearly average age optimal, while the fastest mixing randomized trajectory $\mathbf{P}^{\ast}$ performs slightly better than the Metropolis-Hastings randomized trajectory $\mathbf{P}^{\text{mh}}$.

Theorem~\ref{thm:perform} proved that the fastest mixing randomized trajectory $\mathbf{P}^{\ast}$ is at least factor-$8\mathcal{H}$ optimal. \color{black} Figure~\ref{figure:random_geometric_avg} validates this conclusion: for example, for $n = 90$ ground terminals, the average age for the fastest mixing randomized trajectory $\mathbf{P}^{\ast}$ is approximately a factor $3$ away from the lower bound.

\begin{figure}
\centering
\includegraphics[width=\linewidth]{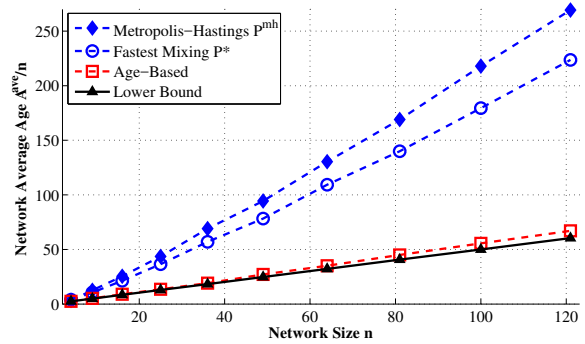}
\caption{Information gathering problem in the Grid graph: network average age as a function of network size $n$ for several proposed trajectories of the mobile agent.}
\label{figure:random_grid}
\end{figure}

\begin{figure}
\centering
\includegraphics[width=\linewidth]{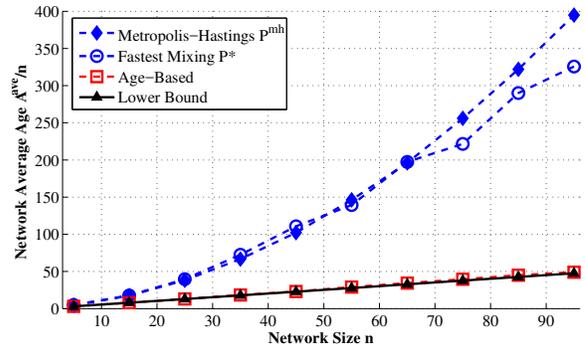}
\caption{Information gathering problem in the Ring graph: network average age as a function of network size $n$ for several proposed trajectories of the mobile agent.}
\label{figure:equal_ring}
\end{figure}

In Figures~\ref{figure:random_grid} and~\ref{figure:equal_ring} we plot the average age performance for several proposed trajectories, as a function of the network size. The age-based policy, again outperforms the two randomized trajectories, and is nearly optimal. We observe that the average age for the fastest mixing randomized trajectory $\mathbf{P}^{\ast}$, namely $A^{\text{ave}}(\mathbf{P}^{\ast})$, is much worse in the ring graph than in the grid graph. This is because the mixing time for the ring graph is much larger than for the grid graph. Similar observation holds in comparing $\mathcal{G}(n, 2/\sqrt{n})$ and the grid graph.

\begin{figure}
\centering
\includegraphics[width=\linewidth]{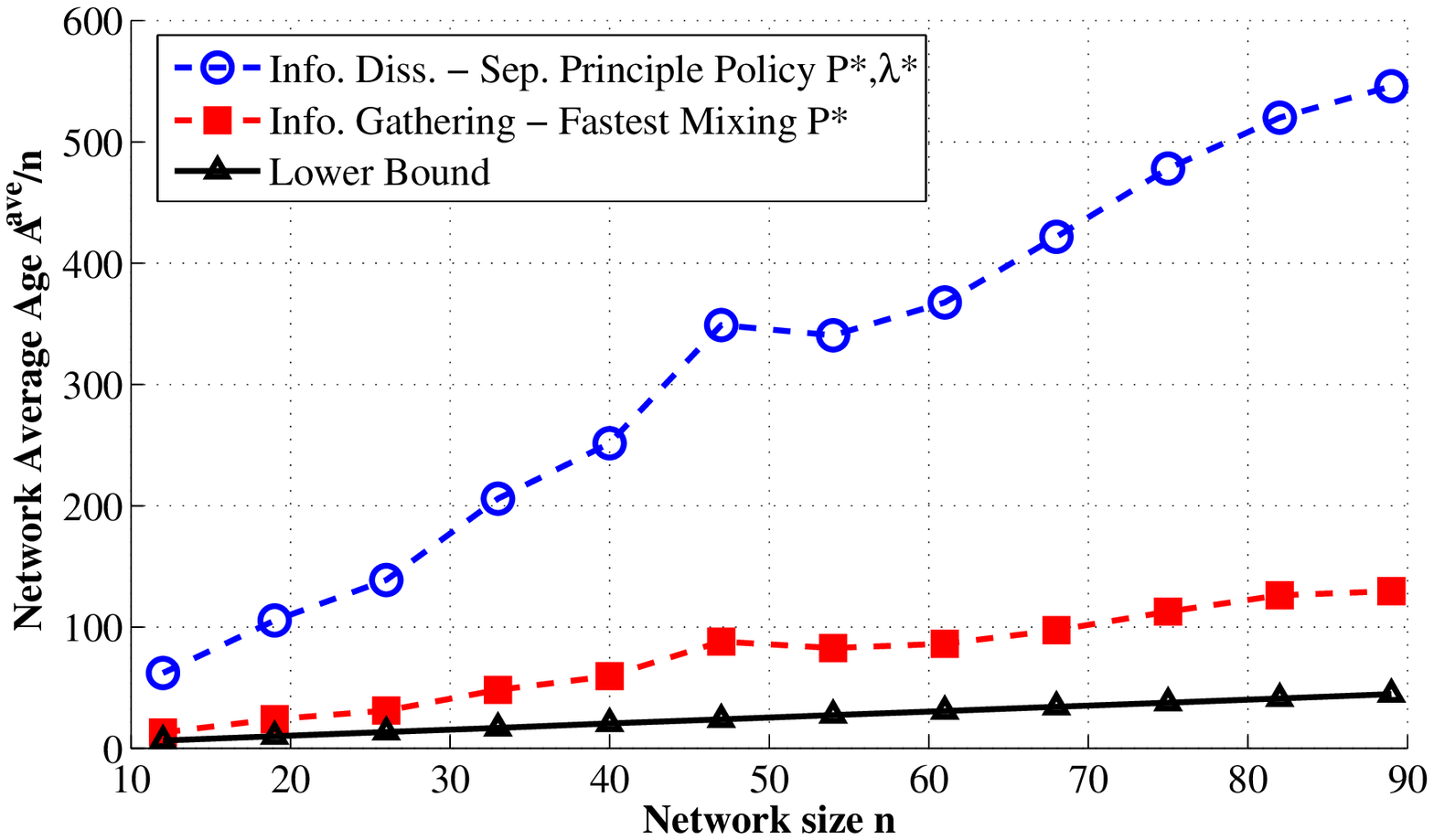}
\caption{Network average age as a function of network size $n$.}
\label{figure:FIFO_Geometric_equal_avg}
\end{figure}

In Figure~\ref{figure:FIFO_Geometric_equal_avg}, we simulate the performance of the separation principle policy for the information dissemination problem, for graph $\mathcal{G}(n,2/\sqrt{n})$, and compare its age performance with the information gathering problem. We observe a significant deterioration of age, as a function of network size $n$, in the information dissemination case in comparison to the information gathering case. This, we note, is the cost of uncontrollable queues in the system on age performance.
%
%
%

\section{Conclusion}
\label{sec:conclusion}

We considered the trajectory planning problem for a mobile agent, that traverses through a mobility graph $G$, to help timely exchange of information updates between a central terminal and a set of ground terminals $V$. 
In the information gathering problem, we showed that a randomized trajectory, namely the fastest-mixing randomized trajectory, is peak age optimal and factor-$\mathcal{H}$ average age optimal. We showed that obtaining an average age optimal trajectory can be NP-hard, while we constricted the peak age optimal trajectory in polynomial time. 
To improve the average age, we proposed an age-based policy, and showed it to be factor-$2$ average age optimal, in a symmetric setting.
In the information dissemination problem, we proposed a separation principle policy, in which the mobile agent follows the fastest mixing randomized trajectory with a simple rate control. We proved that the separation principle policy is factor-$O(\mathcal{H})$ optimal, in both peak and average age.



\bibliographystyle{ieeetr}
\bibliography{bibliography}

\appendix

\subsection{Proof of \thmref{thm:peakage}}
\label{pf:peakage}
Let $A_i^p$ be the peak age for ground terminal $i$. We define $H_{k,i}$ to be the $k\supth$ return time to ground terminal $i$. Then, the $k\supth$ age peak for $A_i(t)$ has a value of $H_{k,i}$. Let $K$ be the total number of returns to $i$ over a time-horizon $T$. Then, the expected peak age of ground terminal $i$ is given by
\begin{equation}
A_i^p = \lim_{T\rightarrow \infty} \mathbb{E}\Bigg[\frac{ \sum\limits_{t = 1}^{t = T} A_i(t)\mathbbm{1}_{\{m(t) = i\}} }{ \sum\limits_{t = 1}^{t = T} \mathbbm{1}_{\{m(t) = i\}} } \Bigg] = \lim_{K\rightarrow \infty} \mathbb{E} \bigg[ \frac{1}{K} \sum\limits_{k = 1}^{t = K}  H_{k,i} \bigg].
\end{equation}

Note that return times to a ground terminal $i$ are i.i.d. random variables given a randomized trajectory $\mathbf{P}$. So, we can use the law of large numbers to get
\begin{equation}
\label{sensorage}
A_i^p = \mathbb{E}[H_{1,i}] = \frac{1}{\pi_i},
\end{equation}
where $\pi_i$ is the stationary distribution for Markov chain $\mathbf{P}$.
The last equality follows from the fact that the expected return time to a state $i$ for an irreducible Markov chain is given by the inverse of its stationary probability. Thus, the network age is given by
\begin{equation}
A^{\text{p}} = \sum_{i \in V} {w_i} A_i^p =\sum_{i \in V} \frac{w_i}{\pi_i}.
\end{equation}

For average age, we define a renewal-reward process using $H_{k,i}$ as our i.i.d. renewal intervals and sum of age $A_i(t)$ during each interval as our reward. Let $T_{k,i} = \sum_{l=1}^{k-1}H_{l,i}$ be the starting time of the $k$th renewal. The total reward in between two visits to ground terminal $i$ is the sum of the $i\supth$ age process $A_i(t)$  across all time-slots during that interval.

Note that, for the $k\supth$ renewal interval, $A_i(t)$ grows from 1 to $H_{k,i}$ over the $H_{k,i}$ time-slots. Thus, the total reward for the $k\supth$ renewal interval is given by -
 \begin{equation}
 \sum_{t=T_{k,i}}^{t=T_{k,i} + H_{k,i}} A_{i}(t) = \sum_{a = 1}^{H_{k,i}} a = \frac{H_{k,i}^2 + H_{k,i}}{2}.
 \end{equation}
Note that this reward is also i.i.d. across renewals as it depends only on $H_{k,i}$. Thus, by application of the elementary renewal theorem for renewal-reward processes we get
\begin{equation}
\label{eq:sensoravgage}
A_i^\text{ave} = \lim_{T\rightarrow \infty}\mathbb{E}\bigg[ \frac{1}{T} \sum_{t = 1}^{t = T} A_i(t) \bigg] =  \frac{\mathbb{E} [H_{1,i}^2 + H_{1,i}] }{2\mathbb{E}[H_{1,i}]}.
\end{equation}
For irreducible Markov chains, we know the following results hold~\cite{aldous2002reversible}:
\begin{equation}
\label{eq:hit1moment}
\mathbb{E}[H_{1,i}] = \frac{1}{\pi_i}, \forall i \in V \textrm{ and}
\end{equation}
\begin{equation}
\label{eq:hit2moment}
\mathbb{E} [H_{1,i}^2] = \frac{-1}{\pi_i} + \frac{2z_{ii}}{\pi_i^2},
\end{equation}
for all $i \in V$, where $z_{ii}$ is the $i\supth$ diagonal element of the matrix $Z = (I - P + \Pi)^{-1}$, with $\Pi$ being a matrix in which all rows are the stationary distribution vector $\pi$: $\Pi_{i,j} = \pi_j$ for all $i,j \in V$.

Substituting \eqref{eq:hit1moment} and \eqref{eq:hit2moment} in  \eqref{eq:sensoravgage}, we get
\begin{equation}
\label{eq:sensoravgage2}
A_i^{\text{ave}} = \frac{z_{ii}}{\pi_i},
\end{equation}
for all $i \in V$, and therefore,
\begin{equation}
A^{\text{ave}} = \sum_{i \in V} {w_i} A_i^\text{ave} =\sum_{i \in V} \frac{w_i z_{ii}}{\pi_i}.
\end{equation}

\subsection{Proof of \thmref{thm:peak_random_tj}}
\label{pf:peakoptmh}
Suppose we could choose any stationary distribution $\pi$ on $V$. Then to minimize the network peak age, we would need to solve the following optimization problem
\begin{align}
\begin{aligned}
\label{statopt}
&\underset{\mathbbm{\pi}}{\text{Minimize}} &&\sum_{i \in V} \frac{w_i}{\pi_i}, \\
&\text{subject to}
&& \sum_{i} \pi_{i} = 1,\\
&&&\pi_{i} \geq 0, \forall i \in V.
\end{aligned}
\end{align}
Using KKT conditions for the optimization problem \eqref{statopt}, it is straightforward to see that
\begin{equation}
\pi_i^* = \frac{\sqrt{w_i}}{\sum\limits_i \sqrt{w_i}}, \forall i \in V.
\end{equation}

Clearly, if we could find a randomized trajectory $\mathbf{P}$ that achieves this stationary distribution $\pi^*$, then it would be peak age optimal. Thus, any randomized trajectory $\mathbf{P}$ that satisfies $\pi^{\ast} = \pi^{\ast}\mathbf{P}$ is peak age optimal. 


%
%

\subsection{Proof of \thmref{thm:peak_opt_all}}
\label{pf:peaklb}
Let $H_{k,i}$ to be the $k\supth$ return time to node $i$. If $K$ is the total number of returns to ground terminal $i$ over a time horizon $T$, then the peak age $A^{\text{p}}_{i}$ is given by 
\begin{equation}
A_{i}^{\text{p}} = \limsup_{T \rightarrow
\infty}\frac{ \sum\limits_{t = 1}^{t = T} A_i(t)\mathbbm{1}_{\{m(t) = i\}} }{ \sum\limits_{t = 1}^{t = T} \mathbbm{1}_{\{m(t) = i\}} } = \limsup_{K \rightarrow \infty}\frac{1}{K}\sum\limits_{k = 1}^{k = K} H_{k,i}.
\end{equation}
Now, the fraction of time-slots in which the mobile agent is at ground terminal $i$, is given by
\begin{equation}
f_i = \lim_{T \rightarrow \infty} \frac{\sum\limits_{t=1}^{t=T}\mathbbm{1}_{\{ m(t) = i\}}}{T} = \lim_{K \rightarrow \infty} \frac{K}{\sum\limits_{k=1}^{k=K} H_{k,i}} = \frac{1}{A_{i}^{\text{p}}},
\end{equation}
and therefore, $A^{\text{p}} = \sum_{i \in V} w_i A^{\text{p}}_{i} = \sum_{i \in V} \frac{w_i}{f_i}$. Note that $f_i$, being the fraction of time-slots the mobile agent is at terminal $i$, is a distribution over $V$. Thus, $A^{\text{p}}$ can be lower bounded by
\begin{equation}
A^{\text{p}} = \sum_{i \in V} w_i A_i^p \geq \min_{\{f_i \geq 0, ~\sum_i f_i = 1 \}}\sum_{i \in V} \frac{w_i}{f_i} = \sum_{i \in V} \frac{w_i}{\pi^*_i},
\end{equation}
where the last equality is obtained by solving the optimization problem, just as in Appendix~\ref{pf:peakoptmh}.

\subsection{Proof of \thmref{thm:equalweight}}
\label{pf:equalweight}

To prove NP-hardness, we establish equivalence between the average age minimization problem and the Hamiltonian cycle problem, in the symmetric setting. We know that more connected the graph, lower is its network average age. Therefore, the average age for G = (V, E) is lower bounded by the average age for the complete graph K(V), given by $\frac{|V|(|V|+1)}{2}$. This lower bound can be obtained by using \thmref{thm:lb} and setting $w_i = 1, ~\forall i$.

If the graph is Hamiltonian, we can achieve this average age lower bound by setting the trajectory equal to a Hamiltonian cycle. This is because in a cyclical trajectory, the agent visits every terminal exactly once in every $|V|$ time-slots. Further, if the graph is not Hamiltonian, the optimal average age is strictly greater than $\frac{|V|(|V|+1)}{2}$. This is because in the absence of a cycle on graph $G$, the agent cannot visit every terminal exactly once every $|V|$ time-slots. Therefore, if an algorithm were to solve the average age problem then the same algorithm could be used to determine whether the graph G is Hamiltonian or not; which is the Hamiltonian cycle problem. Since the Hamiltonian cycle problem is NP-complete, the average age minimization problem must be NP-hard.

\color{black}

\subsection{Proof of \thmref{thm:lb}}
\label{pf:lb}
Let $H_{k,i}$ be the $k\supth$ return time to ground terminal $i$, and $K$ be the total number of returns to $i$ over a time-horizon $T$. Then the average age $A_i^\text{ave}$ is given by (see \apndx{pf:peakage}):
\begin{equation}
\label{lbpfeq1}
A_i^\text{ave} = \lim_{T \rightarrow \infty}\frac{1}{T} \sum\limits_{t = 1}^{t = T} A_i(t) = \lim_{K \rightarrow \infty}\frac{\sum\limits_{k = 1}^{k = K} (H_{k,i}^2 + H_{k,i})}{2\sum\limits_{k = 1}^{k = K}H_{k,i}}.
\end{equation}
Define the empirical first and second moment of return times be $\hat{H}_i \triangleq \frac{1}{K}\sum\limits_{k = 1}^{k = K}H_{k,i}$ and $\hat{H}_{i}^{(2)} \triangleq \frac{1}{K}\sum\limits_{k = 1}^{k = K}H_{k,i}^2$, respectively. Further, define $\hat{\text{Var}}_i \triangleq \hat{H}_{i}^{(2)} - \hat{H}_{i}^{2}$ to be the empirical variance of return times.
From~\eqref{lbpfeq1}, we have
\begin{equation}
A_i^\text{ave} = \frac{1}{2} + \lim_{K \rightarrow \infty} \frac{\hat{H}_{i}^{(2)}}{2\hat{H}_i} = \frac{1}{2} + \lim_{K \rightarrow \infty} \frac{\left( \hat{H}_{i} \right)^{2} + \hat{\text{Var}}_i}{2\hat{H}_i}. \label{eq:n}
\end{equation}
Using Cauchy-Schwarz inequality, we can obtain $\hat{\text{Var}}_i \geq 0$. Applying this to~\eqref{eq:n}, we get
\begin{equation}
\label{lbpfeq2}
A_i^\text{ave} \geq \frac{1}{2} + \lim_{K \rightarrow \infty} \frac{\hat{H}_{i}}{2},
\end{equation}
Let $f_i$ be the fraction of time-slots in which the mobile agent is at  ground terminal $i$. Then,
\begin{equation}
\label{lbpfeq4}
f_i = \lim_{T \rightarrow \infty} \frac{\sum\limits_{t=1}^{t=T}\mathbbm{1}_{\{ m(t) = i\}}}{T} = \lim_{K \rightarrow \infty} \frac{K}{\sum\limits_{k=1}^{k=K} H_{k,i}} = \frac{1}{ \lim_{K \rightarrow \infty} \hat{H}_{i}},
\end{equation}
since $f_i$ is well defined and positive for all trajectories in $\mathbb{T}$. Substituting~\eqref{lbpfeq4} in~\eqref{lbpfeq2} we get $A_i^\text{ave} \geq \frac{1}{2} + \frac{1}{2f_i}$, for all $i$, and
\begin{equation}\label{eq:n2}
A^{\text{ave}} = \sum_{i \in V} w_i A^{\text{ave}}_i \geq \frac{1}{2}\sum_{i \in V} w_i + \frac{1}{2}\sum_{i \in V}\frac{w_i}{f_i}.
\end{equation}
Note that $f_i$, being the fraction of time-slots the mobile agent is at terminal $i$, is a distribution over $V$. Thus, the average age in~\eqref{eq:n2} can be lower bounded by
\begin{align}
A^{\text{ave}} &\geq \frac{1}{2}\sum_{i \in V}w_i + \frac{1}{2}\min_{ \{f_i \geq 0, ~\sum_i f_i = 1 \} } \sum_{i \in V}\frac{w_i}{f_i}, \nonumber  \\
&= \frac{1}{2}\sum_{i \in V}w_i + \frac{1}{2} \sum_{i \in V}\frac{w_i}{\pi^{\ast}_i}, \nonumber 
\end{align}
which proves the result.

\subsection{Proof of \thmref{thm:convex_avg_age}}
\label{pf:convex_avg_age}
From \cite{boyd2004fastest}, we know that the fastest mixing, reversible Markov chain on a graph $G(V,E)$ having the stationary distribution $\pi$ can be found by formulating the following convex program:
\begin{align}
\begin{aligned}
\label{op:boyd}
&\underset{\mathbf{P}}{\text{Minimize}} && || D^{1/2}\mathbf{P}D^{-1/2} - qq^T||_2, \\
&\text{subject to}
&& P_{i,j} \geq 0, ~\forall (i,j)\\
&&&\mathbf{P}\mathbf{1} = \mathbf{1},\\
&&& \pi^* \mathbf{P} = \mathbf{P}^T \pi^*,\\
&&& P_{i,j} = 0, \forall (i,j) \notin E.
\end{aligned}
\end{align}
Here $D = \text{diag}(\pi^*)$ and $q = (\sqrt{\pi_1^*},\sqrt{\pi_2^*},...,\sqrt{\pi_n^*}).$ Note that we do not require reversibility, so we can replace the detailed balance constraint $\pi^* \mathbf{P} = \mathbf{P}^T \pi^*$ with the global balance constraint $\pi^* \mathbf{P} = \pi^*$. Also, left and right multiplying $(D^{1/2}\mathbf{P}D^{-1/2} - qq^T)$ by matrices $D^{-1/2}$ and $D^{1/2}$, respectively, does not change the spectral norm; since $\mathbf{P}$ has the same eigen-values as $D^{1/2}\mathbf{P}D^{-1/2}$ and  $qq^T$ has the same eigen-values as $D^{-1/2}qq^T D^{1/2}$ \cite{boyd2004fastest}. Further, observe that $D^{-1/2}qq^TD^{1/2} = qq^T = \Pi^*$, where $\Pi_{i,j}^{\ast} = \pi_i^{\ast}~\forall~i,j \in V.$ Thus, the optimization problem reduces to~\eqref{op:avgh}. This proves the required result.

\subsection{Proof of \thmref{thm:perform}}
\label{pf:perform}
Note that the peak age for the fastest-mixing randomized trajectory $\mathbf{P}^{\ast}$ is given by $A^{\text{p}}(\mathbf{P}^{\ast}) = \sum_{i \in V} \frac{w_i}{\pi^{\ast}_i}$,
since $\pi^{\ast}\mathbf{P}^{\ast} = \pi^{\ast}$. From Theorem~\ref{thm:lb}, a lower bound on average age is given by
\begin{equation}
\label{eq:mm1}
A^{\text{ave}}_{\text{LB}} = \sum_{i \in V}\frac{1}{2}\left( \frac{w_i}{\pi^{\ast}_i} + w_i \right) = \frac{1}{2}A^{\text{p}}(\mathbf{P}^{\ast}) + \frac{1}{2}\sum_{i \in V}w_i.
\end{equation}

To prove the result, it suffuses to argue that $A^{\text{ave}}(\mathbf{P}^{\ast})/A_{\text{LB}} \leq 8\mathcal{H}$. From~\eqref{eq:mm1} and \lemref{lem:approxlem}, we get
\begin{align}
\frac{ A^{\text{ave}}(\mathbf{P}^{\ast}) }{ A^{\text{ave}}_{\text{LB}} } &\leq \frac{ 4\mathcal{H}A^{\text{p}}(\mathbf{P}^{\ast}) + \sum_{i \in V}w_i }{ \frac{1}{2}A^{\text{p}}(\mathbf{P}^{\ast}) + \frac{1}{2}\sum_{i \in V}w_i }, \\
&\leq 8\mathcal{H},
\end{align}
since $\mathcal{H}$ is always greater than or equal to $1$.

\subsection{Proof of Theorem~\ref{thm:age-based}}
\label{pf:thm:age-based}

The number of steps taken to cover every vertex of a graph by performing a depth first search (DFS) traversal is upper bounded by $2|V|$, since every vertex is visited at least once and the sum total of visits after the first visit to all nodes is at most $|V|$. This is because every repeated visit to a vertex means that at least one new vertex was visited. Thus, every location gets visited at least once in every $2|V|$ time-slots. This implies that the average age of every terminal can be upper bounded by $\frac{(2|V|+1)}{2}$.

However, from our earlier discussion, we know that the average age of any terminal is lower bounded by $\frac{(|V|+1)}{2}$ if all the weights are $1$. Combining the upper and lower bounds, we have the required result.

\color{black}

\subsection{Proof of \lemref{lem:buff1}}
\label{pf:buff1}

\textbf{Derivation of System Time}
The proof is a discretized version of the proof for M/G/1 queues with vacations using residual service times as discussed in \cite{bertsekas1992data}.

Let us define the residual service time for an update at time $t$, given by $R(t)$, as the amount of time remaining until the update currently at the head of the queue is complete, excluding the current time-slot. If the queue is empty, $R(t)$ equals zero.

From \cite{bertsekas1992data} we know that the expected waiting time in the queue can be found using the residual service times as follows
\begin{equation}
 \label{eq:vacation}
 \EX{T_Q} = \frac{\EX{R}}{1-\rho},
 \end{equation}
  where $\rho = \frac{\lambda}{\mu}$, $\EX{S} = \frac{1}{\mu}$ and $\EX{R} = \lim\limits_{T \rightarrow \infty} \mathbb{E} \bigg[\frac{1}{T}  \sum\limits_{t=0}^{t = T} R(t)\bigg]$. As in \cite{bertsekas1992data}, $\EX{R}$ can be computed using a graphical argument. Let service times for the $m$th packet be $X_m$, and let the $k$th vacation time be $V_k$. Let the total number of packets served be $M(T)$ and the total number of vacations be $L(T)$, over the entire time-horizon $T$. Then, we have

  \begin{multline}
  \frac{1}{T}  \sum\limits_{t=0}^{t = T} R(t) = \frac{1}{2} \frac{M(T)}{T} \frac{\sum\limits_{m=1}^{m = M(T)} (X_m^2 - X_m)}{M(T)} \\
  + \frac{1}{2} \frac{L(T)}{T} \frac{\sum\limits_{k=1}^{k = L(T)} (V_k^2 - V_k)}{L(T)}.
  \end{multline}
  Using the strong law of large numbers and the fact that $\frac{M(T)}{T} \rightarrow \lambda$ and $\frac{L(T)}{T} \rightarrow \frac{(1-\rho)}{\EX{V}}$, we get

  \begin{equation}
  \label{eq:remain}
  \EX{R} = \frac{\lambda (\EX{S^2} - \EX{S})}{2} + \frac{(1-\rho)(\EX{V^2} - \EX{V})}{2\EX{V}}.
  \end{equation}
 Combining \eqref{eq:vacation} and \eqref{eq:remain}, we get

 \begin{equation}
 \EX{T_Q} = \frac{\lambda \mathbb{E}[S^2]-\rho}{2(1-\rho)} + \frac{\EX{V^2}}{2\EX{V}}-\frac{1}{2}.
 \end{equation}

The total time spent in the system by a packet is given by the sum of its waiting time in the queue and its processing time, which implies
 \begin{equation}
 \EX{T} = \EX{S + T_Q}=\frac{1}{\mu} + \frac{\lambda \mathbb{E}[S^2]-\rho}{2(1-\rho)} + \frac{\EX{V^2}}{2\EX{V}}-\frac{1}{2},
 \end{equation}
since $\EX{S} = \frac{1}{\mu}$.

\textbf{Average Age}: Consider a $Ber/G/1$ queue with vacations that has i.i.d. packet inter-arrival times $X_1, X_2, ...$ Let $T_n$ be the total time spent in the system by the $n\supth$ packet. Then, the average age is given by~\cite{kaul2012real}:
\begin{equation}
A^\text{ave} = \frac{1}{\lambda} + \lambda \mathbb{E}[X_n T_n],
\end{equation}
where $\frac{1}{\lambda} = \mathbb{E}[X_n].$ To evaluate the term $\mathbb{E}[X_n T_n]$, we observe that larger inter-arrival times $X_n$ between packets mean lesser wait times in the system $T_n$ for individual packets. Thus, $X_n$ and $T_n$ are negatively correlated. Note that for negatively correlated random variables the following holds
\begin{equation}
\EX{X_n T_n} \leq \EX{X_n} \EX{T_n}.
\end{equation}
This implies
\begin{equation}
A^\text{ave} \leq \frac{1}{\lambda} + \lambda \mathbb{E}[X_n] \EX{T_n} = \EX{X_n} + \EX{T_n} = A^\text{p},
\end{equation}
since $\mathbb{E}[X_n] = 1/\lambda$.

\subsection{Proof of \lemref{lem:buff2}}
\label{pf:buff2}
Consider a randomized trajectory $\mathbf{P}$ and Bernoulli arrival rates $\bm{\lambda} = (\lambda_1, \lambda_2, \ldots )$. From the arguments made in Section~\ref{sec:info_diss}, we know that the peak age for the ground terminal $i$ is upper-bounded by the peak age of a discrete time FCFS Ber/G/1 queue with vacations, for which the service times and vacation times have the same distribution as the inter-visit times $H_{1,i}$.
Applying~\lemref{lem:buff1} we obtain
\begin{equation}
\label{eq:peakbuff1}
A^{\text{p}}_{i} \leq \frac{1}{\pi_i}\left[ 1 + z_{ii} +\frac{1}{\rho_i} + \frac{z_{ii}\rho_i}{1 - \rho_i}\right] - \frac{\rho_i}{1-\rho_i} - 1 \triangleq A^{\text{UB}}_{i},
\end{equation}
where we have used the first and second moment of inter-visit times $H_{1,i}$~\cite{aldous2002reversible}:
\begin{equation}
\label{eq:hit1moment1}
\mathbb{E}[H_{1,i}] = \frac{1}{\pi_i}, ~
\mathbb{E} [H_{1,i}^2] = \frac{-1}{\pi_i} + \frac{2z_{ii}}{\pi_i^2}, \forall i \in V.
\end{equation}

Similarly, we know that the average age for the ground terminal $i$ is also upper-bounded by the average age for the FCFS Ber/G/1 queue with vacations. Using the fact that $A^{\text{ave}} \leq A^{\text{p}}$ for the Ber/G/1 queue with vacations (see Lemma~\ref{lem:buff1}), we get $A^{\text{ave}}_i \leq A^{\text{UB}}_{i}$.

\subsection{Proof of \thmref{thm:buff_perform}}
\label{pf:buff_perform}
We want to solve the upper bound age minimization problem, which can be stated as:
\begin{align}
\begin{aligned}
&\underset{\mathbf{P}, \bm{\rho}}{\text{Minimize}} &&\sum_{i \in V} w_i A_i^\text{UB}, \\
&\text{subject to} && P_{i,j} \geq 0, ~\forall (i,j), \\
&&&\mathbf{P}\mathbf{1} = \mathbf{1},\\
&&& P_{i,j} = 0, ~\forall (i,j) \notin E,\\
&&& \mathbf{P}\text{ is irreducible.}
\end{aligned}
\end{align}
We first find the optimal packet generation rates given a random walk $\mathbf{P}$. Observe that the optimal queue utilization factors $\rho_i$ can be solved for given any fixed irreducible random walk $\mathbf{P}$, i.e.
\begin{equation}
\rho_i^*(\mathbf{P}) = \text{arg}\min_{\rho_i \in [0,1]} A_i^\text{UB}(\mathbf{P}, \rho_i) = \frac{1}{1+\sqrt{z_{ii}-\pi_i}}
\end{equation} and
\begin{equation}
\min_{\rho_i \in [0,1]} A_i^\text{UB}(\mathbf{P}, \rho_i) = A_i^\text{UB}(\mathbf{P},\rho_i^*) = \frac{z_{ii} - \pi_i + 2\sqrt{z_{ii}-\pi_i}+2}{\pi_i}.
\end{equation}
Thus, the upper bound age minimization problem reduces to
\begin{align}
\begin{aligned}
&\underset{\mathbf{P}}{\text{Minimize}} &&\sum_{i \in V} w_i\bigg( \frac{z_{ii} - \pi_i + 2\sqrt{z_{ii}-\pi_i}+2}{\pi_i}\bigg), \\
&\text{subject to} && P_{i,j} \geq 0, ~\forall (i,j), \\
&&&\mathbf{P}\mathbf{1} = \mathbf{1},\\
&&& P_{i,j} = 0, ~\forall (i,j) \notin E,\\
&&& \mathbf{P}\text{ is irreducible.}
\end{aligned}
\end{align}

Now, we can relate the network age upper bound, given a random walk $\mathbf{P}$, to its mixing time $\mathcal{H}$. We assume optimal packet generation rates ${\rho_i^*}(\mathbf{P})$.
\begin{align}
\sum_{i \in V} w_i A_i^\text{UB}(P,\rho_i^*(P)) &=
\sum_{i \in V} w_i\bigg( \frac{z_{ii} - \pi_i + 2\sqrt{z_{ii}-\pi_i}+2}{\pi_i}\bigg), \nonumber \\
&\leq \sum_{i \in V} w_i\bigg(\frac{\mathcal{Z} + 2\sqrt{\mathcal{Z}}+2}{\pi_i}\bigg), \nonumber \\  
&\leq \sum_{i \in V} w_i\bigg(\frac{4\mathcal{H} + 4 \sqrt{\mathcal{H}} + 2}{\pi_i}\bigg), \nonumber 
\end{align}
where inequalities follow from the same argument as in the proof of \lemref{lem:approxlem}. 
Setting $\mathbf{P} = \mathbf{P}^*$, we obtain
\begin{equation}
\sum_{i \in V} w_i A_i^\text{UB}(\mathbf{P}^*,\rho_i^*(\mathbf{P}^*)) \leq \sum_{i \in V} w_i\bigg(\frac{4\mathcal{H} + 4 \sqrt{\mathcal{H}} + 2}{\pi_i^*}\bigg),
\end{equation}
where $\mathcal{H}$ is the mixing time of $P^*$. Note that $\sum_{i \in V}\frac{w_i}{\pi^{\ast}_i}$ is the optimal peak age in the information gathering problem, i.e. $A^{\text{p}\ast}_{\mathcal{G}} = \sum_{i \in V}\frac{w_i}{\pi^{\ast}_i}$. This gives,
\begin{equation}
\frac{A^\text{UB}(\mathbf{P}^*,\bm{\rho}^*)}{A^{\text{p}\ast}_{\mathcal{G}}} \leq 4\mathcal{H} + 4 \sqrt{\mathcal{H}} + 2. \label{eq:no1}
\end{equation}
Due to the presence of queues we have $A^{\text{p}\ast}_{\mathcal{G}} \leq A^{\text{p}\ast}_{\mathcal{D}}$. This,~\eqref{eq:no1}, and the fact that  $A^\text{p}(\mathbf{P}^*,\bm{\rho}^*) \leq A^\text{UB}(\mathbf{P}^*,\bm{\rho}^*)$, yields the peak age bound on the separation principle policy:
\begin{equation}
\frac{A^\text{p}(\mathbf{P}^*,\lambda^*)}{A^{\text{p}\ast}_{\mathcal{D}}} \leq 4\mathcal{H} + 4 \sqrt{\mathcal{H}} + 2, \nonumber 
\end{equation}
since $\bm{\rho}^* = \bm{\lambda}^{\ast}$.

From the discussion following Theorem~\ref{thm:lb}, we know that $2A^{\text{ave}\ast}_{\mathcal{G}} \geq A^{\text{p}\ast}_{\mathcal{D}}$. Also, $A^{\text{ave}\ast}_{\mathcal{G}} \leq A^{\text{ave}\ast}_{\mathcal{D}}$ and $A^\text{ave}(\mathbf{P}^*,\bm{\rho}^*) \leq A^\text{UB}(\mathbf{P}^*,\bm{\rho}^*)$. Combining these with~\eqref{eq:no1} gives us
\begin{equation}
\frac{A^\text{ave}(\mathbf{P}^*,\lambda^*)}{A^{\text{ave}\ast}_{\mathcal{D}}} \leq 8\mathcal{H} + 8 \sqrt{\mathcal{H}} + 4, \label{eq:no2}
\end{equation}
since $\bm{\rho}^* = \bm{\lambda}^{\ast}$.

\end{document}